\documentclass[twocolumn]{aastex7}
\usepackage[]{units}
\usepackage[update,prepend]{epstopdf}
\usepackage{graphicx}
\usepackage{amssymb}
\usepackage{amsmath}
\usepackage{threeparttable}
\usepackage{hyperref}
\usepackage{color}
\usepackage{gensymb}
\usepackage{mathtools}

\usepackage{makecell}
\usepackage{bm}

\newcommand{\kms} {\,km\,s$^{-1}$}

\newcommand{\Msun}{\,M$_\odot$}

\mathchardef\mhyphen="2D
\shorttitle{White dwarfs in wide binaries}
\shortauthors{}

\graphicspath {{figs/} }

\begin{document}

\title{White dwarfs in wide binaries: the strong effects of stellar evolution and mass loss}

\author[0000-0003-4250-4437]{Hsiang-Chih Hwang}
\affiliation{Institute for Advanced Study, Einstein Dr., Princeton NJ 08540}
\email{hsiangchih.hwang@gmail.com}

\author[0000-0001-6100-6869]{Nadia L. Zakamska}
\affiliation{Department of Physics \& Astronomy, Johns Hopkins University, Baltimore, MD 21218, USA}
\affiliation{Institute for Advanced Study, Einstein Dr., Princeton NJ 08540}
\email{zakamska@jhu.edu}

\begin{abstract}
We examine the statistics of main-sequence / main-sequence, main-sequence / white-dwarf and white-dwarf / white-dwarf wide binaries at $10^{2.5}-10^4$ AU separations in {\it Gaia} data. For binaries containing a white dwarf, we find a complex dependence of the wide binary fraction on the white dwarf mass, including a steep decline as a function of mass at $>0.6$\Msun. Furthermore, we find that wide binaries containing white dwarfs have significantly lower eccentricities than main-sequence binaries at the same separations. To model these observations, we compute the effects of post-main-sequence mass loss on the orbital parameters of wide binaries in all regimes of timescales, from secular to impulsive, and incorporate this dynamics in a population synthesis model. We find that adiabatic expansion of the orbits in binaries with slow enough evolutionary processes is the most likely explanation for the puzzling eccentricity distribution of white dwarf wide binaries. The steeply declining white dwarf binary fraction as a function of mass requires that the timescale for mass loss must be significantly shorter for high-mass stars ($10^3-10^4$ years) than for the low-mass ones. We confirm previous studies that suggested that recoil in the range $0.25-4$ km s$^{-1}$ is required to explain the observed distribution of separations of white dwarf wide binaries. Finally, for low-mass white dwarfs ($<0.5$\Msun), we see interesting signatures of their formation due to close binary evolution in their wide binary statistics. Our observations and modeling provide a novel dynamical constraint on the mass-loss stages of stellar evolution that are difficult to probe with direct observations.
 
\end{abstract}
%\keywords{binaries: general --- stars: evolution -- stars: mass-loss --- white dwarfs}

\section{Introduction}

Conducting a thorough census of stellar binaries and higher-order multiples remains a major goal of modern astrophysical surveys. The observational challenge for a complete survey and a robust measurement of the multiplicity lies in the necessity to employ different techniques for different orbital separations and mass ratios, from radial velocities, to astrometric measurements or the orbital motion, to identification of comoving companions \citep{abt76, toko06, ragh10, toko14, klei17, Moe2017}. In the last few years, {\it Gaia} has made it possible to identify wide binaries and hierarchical multiples in large numbers outside of the local Solar neighborhood \citep{El-Badry2018,Hartman2020,bran21,hwan21wide,El-Badry2021,Fezenko2022,halb23}. 

Binary fraction depends on component masses, separation, metallicity and age, often in a complicated and even non-monotonic fashion \citep{Moe2017, hwan21wide}. These measurements provide insights into binary formation and evolution processes. For example, thanks to its precision distances and proper motions, {\it Gaia} has enabled statistical measurements of the eccentricities of well-separated binaries. This measurement by \citet{Hwang2022ecc} demonstrated that there is a significant dynamical difference between binaries at separations $<10^2$ AU whose eccentricities are drawn from a uniform distribution and those with separations $>10^2$ AU whose eccentricities are typically significantly higher. Thus it is now possible to clearly define the boundary separating `close' and `wide' binaries purely by their dynamical conditions, which likely reflect different formation mechanisms \citep{Hwang2022ecc, Hwang2022twin}.

The effects of metallicity and age are particularly difficult to disentangle since metal-poor populations in the Galaxy also tend to be the oldest \citep{hwan21wide}. The disambiguation can be performed with the help of statistical `kinematic ages' which can be measured using {\it Gaia} proper motions, thanks to the overall increase of the velocity dispersion of a given stellar population as it ages \citep{hame19, chen19, hwan20b}. 

Binaries at separations $<10^4$ AU are not easily disrupted by passages of other stars, molecular clouds, Galactic tides or other substructure \citep{jian10, hami22, moda23, Hamilton2024}. Therefore, the distribution of such binaries likely reflects the complex star formation processes \citep{xu23}, which may be dependent on metallicity and separation, plus on the subsequent orbital evolution due to any internal processes. Close binary companions ($<10^2$ AU) affect each other's stellar evolution and their common orbit in profound ways \citep{pacz71}. For wide binaries ($>10^2$ AU) not affected by mass transfer, the components evolve largely independently, and the most important process that potentially has an impact on the orbital evolution is the stellar mass loss that most stars undergo on their way from the main sequence to a white dwarf and the closely related velocity recoil, or kick, that may be imparted to the white dwarf during its formation. This process is the focus of this work. 

In this paper we measure the wide-binary fraction of white dwarfs and the orbital eccentricity of wide binaries containing white dwarfs. We further present theoretical models for the orbital evolution of wide binaries resulting from the significant mass loss the stars undergo as they evolve from the main sequence into white dwarfs. In Section \ref{sec:data}, we describe the sample selection, measurement methods and key observational results. In Section \ref{sec:model}, we present the binary population synthesis model that incorporates orbital evolution due to mass loss and recoil during post-main-sequence evolution. In Section \ref{sec:results}, we present the results of the simulation and discuss model trends, and we summarize in Section \ref{sec:conclusions}. 

We define the binary fraction as 
\begin{equation}
  f_b=N_b/N_{\rm all},  \label{eq:bfraction}
\end{equation}
where $N_b$ is the number of stars in binaries and $N_{\rm all}$ is the total number of stars. This is different from some of the definitions of the multiplicity fractions defined as the fraction of the {\it systems} in the population that are multiples (e.g., eq. 1 of \citealt{dona23}). For example, if the survey contains 300 stars, of which 200 are in 100 binaries and 100 more are singles, our definition would yield a binary fraction of $2/3$, whereas the definition of \citet{dona23} would yield $1/2$ because half of the systems are multiples. Our rationale for adopting this somewhat less common definition is that it is significantly easier to self-consistently define the binary fraction as a function of mass, which is a parameter that describes the stars in the multiple system separately. We convert the values from the literature to our definition when necessary. 

\section{Sample selection and measurements}
\label{sec:data}

%\subsection{White dwarf selection}
\subsection{Sample selection}
\label{sec:selection}

In this paper, we focus on the wide binary fractions of white dwarfs and main-sequence stars. We select white dwarfs using {\it Gaia} EDR3 white dwarf catalog by \citet{gent21} who tabulate $P_{\rm WD}$, the probability of an object to be a white dwarf. We select high-confidence white dwarfs using the selection criterion $P_{\rm WD}>0.9$. We further use the {\it Gaia} EDR3 wide binary catalog by \citet{El-Badry2021}, where wide binaries are identified by having parallaxes and proper motions consistent with a gravitationally bound system. The all-sky catalog contains 2 million wide binaries out to 1 kpc from the Sun. The separations of wide binaries range from $10^2$\,AU to $\sim10^5$\,AU. Wide binary candidates with wider separations at $>10^4$\,AU have a higher chance to be a chance-alignment pair and they are more likely to be affected by post-birth encounters and Galactic tides. Therefore, we focus on wide binaries at $<10^4$\,AU, where the contamination of chance-alignment pairs is negligible ($<0.1$ percent).  

Following the {\it Gaia} wide binary catalog \citep{El-Badry2021}, we adopt the selection criteria of \texttt{parallax\_over\_error}$>5$, \texttt{parallax\_error}$<2$, the number of neighbors $\le 30$ to avoid clustered regions \citep{El-Badry2021}. To have reliable photometry in {\it Gaia}, we require the sample to have fluxes over error $>10$ in G, BP, and RP bands, and \texttt{phot\_bp\_rp\_excess\_factor}$<1.8$ to avoid the downgraded BP and RP photometry in the crowded field \citep{Evans2018}. The criterion \texttt{ruwe}$<1.4$ is used to ensure the quality of astrometric measurements \citep{Lindegren2018}. We limit our sample to have Galactic latitudes $|b|>10$\,deg to avoid the dusty think disk, and we do not explicitly correct for extinction. 

We focus on the sample where parallaxes $>2$\,mas, i.e. distances $<500$\,pc. With the astrometric and BP/RP photometry criteria above, {\it Gaia}'s limiting G-band magnitude is $\sim19$\,mag, corresponding to the absolute magnitude of $M_G=10.5$\,mag. Hence, we define the main-sequence wide binary fraction to be the fraction of a target sample (e.g. white dwarfs) that have wide main-sequence companions with $M_G<10.5$\,mag and BP-RP$>0.8$ mag (masses $0.433-1.066$\,\Msun). The color criterion is introduced in order to avoid young bright main-sequence stars, whose numbers depend sensitively on the recent star-formation history of the Galaxy, but none of the results qualitatively change if this criterion is relaxed or removed. The wide binary fractions in this paper only consider those with projected separations between $10^3$ and $10^4$\,AU. With the parallax criterion of $>2$\,mas, these wide binaries have angular separations of $>2$\,arcsec, ensuring the high completeness of wide binaries \citep{El-Badry2021} and reliable BP/RP photometry \citep{Evans2018}. 

These color and magnitude selection criteria apply only to the main-sequence companions to target stars in our sample. Many WDs are significantly fainter than $M_G=10.5$\,mag, and we are not claiming to be complete in WD selection within 500 pc. However, once a population of target WDs (or any other stars) within 500 pc is identified, our sample is complete to main-sequence (MS) companions meeting criteria above. Therefore, the wide binary fraction in this paper is defined in a somewhat narrow way, with strong restrictions on the binary separations and on the companion properties, but it can be applied self-consistently in the data to any population of stars and in the stellar population model.

\subsection{Mass derivations}
\label{sec:mass}

We derive the masses for white dwarfs using their locations in the H-R diagram. Since $\sim80$\% of white dwarfs are hydrogen-atmosphere white dwarfs \citep{GentileFusillo2019}, we use the hydrogen-atmosphere model when deriving the mass. We do not find a significant difference in wide binary fractions between hydrogen-atmosphere (DA) and helium-atmosphere (DB) white dwarfs. Following the implementation of \citet{chen19}, we use the low-mass ($\lesssim0.5$\,\Msun) model from \citet{font01}, the middle-mass (about $0.5$-$1$\,\Msun) model from \cite{rene10}, and the high-mass model ($>1$\,\Msun) with O/Ne cores from \cite{Camisassa2019}. The typical uncertainties of this procedure for white dwarfs with $G=18-20$ mag (the majority of objects in \citealt{gent21}) are $0.15-0.2$ \Msun\ \citep{Crumpler2024}, but that turns out to be sufficient for our purposes of detecting the mass dependence of the white dwarf binary fraction. To calculate the mass of the progenitor from the mass of the observed white dwarf, we use the initial-to-final mass relation from \cite{Cummings2018}. 

The main-sequence progenitors of low-mass white dwarfs with masses $<0.5$\,\Msun\ have a theoretical main-sequence lifetime longer than the Hubble time, and it is generally believed that these low-mass white dwarfs are the consequence of close binary evolution. The precise white dwarf mass cutoff below which single stellar evolution cannot produce a white dwarf over the Hubble time is not well-known and depends sensitively on the initial-to-final mass relationship, which in turn likely depends on the metallicity. For the \cite{Cummings2018} initial-to-final mass relationship, this mass is 0.56\Msun. Therefore, most or all of the white dwarfs whose masses are measured to be $\la 0.5$\,\Msun\ are in fact in close binaries spatially unresolved by our observations. One possibility is that the close companion is  another white dwarf, more massive and therefore less luminous \citep{Iben1990, Brown2011,Brown2016}. Another possibility is a red, low-mass main-sequence companion or a brown dwarf, invisible in the optical photometry \citep{Maxted2006, Kruckow2021, VanRoestel2021}. 

For main-sequence stars, which are brighter, the uncertainties are significantly lower. \cite{Hwang2024} measure dynamical masses across the Hertzsprung-Russell diagram using the orbital motion of wide binaries and provide a fitting for main-sequence mass with a 15-order polynomial, which may be subject to numerical precision error. Here we use the 5-order polynomial to map from BP-RP colors (in mag) to the single-star main-sequence mass (in \Msun), with the fitting parameters from the lowest to highest order: $1.9577$, $-1.4539$, $0.4208$, $0.0281$, $-0.0344$, $0.0042$. This fitting is valid for input BP-RP color between 0 and 4 mag, corresponding to main-sequence masses of 1.96 and 0.21\,\Msun, respectively. The typical uncertainty is $\sim0.08$\,\Msun\ for main-sequence single stars. 

\subsection{Mass-dependent white dwarf binary fraction}

\begin{figure}
	\centering
	\includegraphics[width=1\linewidth]{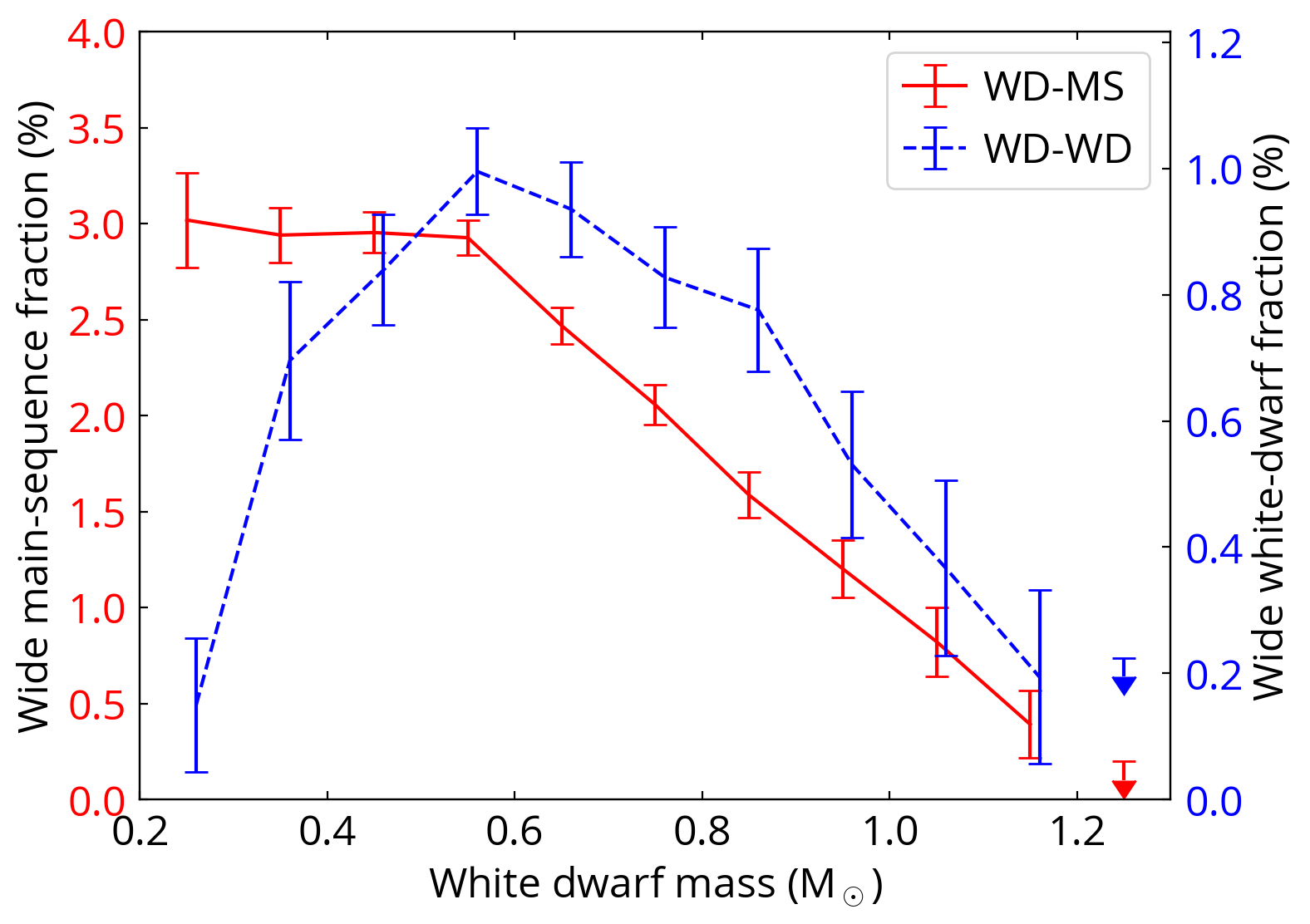}
	\caption{White dwarfs' wide binary fractions as a function of white dwarf masses. Above $0.5$\,\Msun\ white dwarf masses, the trends of the two lines are similar, reflecting a nearly constant main-sequence-to-white-dwarf ratio in the wide companions at different white dwarf masses. The two lines show drastically different trends at $<0.5$\,\Msun, likely due to the nature of the low-mass white dwarfs. }
	\label{fig:wbf-mass}
\end{figure}

Following eq. (\ref{eq:bfraction}), we define the mass-dependent white dwarf wide binary fraction as
\begin{equation}
    f_{\rm WD-MS} = \frac{N_{\rm WD-MS}}{N_{\rm WD}}.
    \label{eq:frac2}
\end{equation}
Here $N_{\rm WD-MS}$ is the number of white dwarfs of a given mass that are in wide binaries with main sequence stars and $N_{\rm WD}$ is the total number of white dwarfs of that mass in the survey. As noted above, the binary fraction is defined with strict criteria on the MS companion (separations $10^3-10^4$ AU, masses $0.433-1.066$\,\Msun) where our observational selection techniques are complete. Even if the survey volume depends on the properties of the target WD (e.g., more massive fainter WDs are detectable to a smaller distance), the same WDs appear in the numerator and in the denominator of the definition, allowing for a robust binary fraction determination. With this definition, Fig.~\ref{fig:wbf-mass} presents the wide binary fraction as a function of white dwarf masses. The red line shows the fraction among the white dwarfs with given masses that have a main-sequence wide companion (hereafter WD-MS binaries).

The blue line represents the white-dwarf / white-dwarf (WD-WD) binary fraction defined by analogy with the equation above. For every WD in the sample, we ask whether it is in a $10^3-10^4$ AU binary with another WD. Therefore, every WD-WD binary makes two appearances in the blue line in Fig.~\ref{fig:wbf-mass}: one time in the mass bin corresponding to the lower mass companion and another time in the mass bin corresponding to the higher mass companion (or twice in the same bin if the companion masses are the same). The error bars are the 1-sigma Poisson uncertainties. 

We see that as a function of white dwarf mass, the WD-MS wide binary fraction starts with a plateau at $f_{\rm WD-MS}=3\%$ for $m_{\rm WD}=0.2-0.6$\,\Msun\ and then steeply declines by a factor of $\sim 6$ at white dwarf masses $m_{\rm WD}=0.6-1.2$\,\Msun. As discussed in more detail in the next section, this mass dependence is in stark contrast to that of the MS-MS wide binary fraction, which increases with the MS mass \citep{Moe2017}. The WD-WD binary fraction mimics the same trend at high WD masses: it declines from $f_{\rm WD-WD}=0.6\%$ to 0.1\%. However, at low WD masses the mass dependence of $f_{\rm WD-WD}$ is qualitatively different from that of $f_{\rm WD-MS}$. There is no plateau, and instead the binary fraction rises steeply for $m_{\rm WD}=0.2-0.6$\,\Msun. 

Since the main-sequence stars are significantly brighter than white dwarfs, if the white dwarf is detected in a survey then we are unlikely to miss its main-sequence wide binary companion. Therefore, in a given bin of white dwarf mass, both the numerator and the denominator of eq. (\ref{eq:frac2}) probe the same volume. But for the WD-WD binary fraction shown in Fig.~\ref{fig:wbf-mass}, one possible concern is that we may be missing some wide binary white dwarf companions of the low-mass white dwarfs: because more massive white dwarfs have a smaller radius, they are fainter. This effect could artificially lower the WD-WD binary fraction of low-mass white dwarfs. To investigate how the completeness of our sample affects the measured WD-WD binary fraction, we have experimented with different parallax cuts -- parallax$>$2, 3, 4, and 5 mas. The results are all qualitatively similar to those shown in Fig.~\ref{fig:wbf-mass}, which is made with the parallax$>$4 mas cut (which is more conservative than our primary selection $>$2 mas described in Section \ref{sec:selection}), except for the sample with the $>$5 mas cut where the WD-WD binary fraction is rather flat on the low-mass end, resembling the red line. However, at that point the number of objects is so low that the Poisson error bars are very high, and therefore in what follows we take the relationship shown in Fig.~\ref{fig:wbf-mass} at face value. 

It has long been known that white dwarfs with masses $\la 0.5$\,\Msun\ must originate from close binary evolution \citep{pacz71} because for single stars which could evolve into white dwarfs of such low masses the MS lifetimes of their MS progenitors are longer than the Hubble time, although a direct confirmation of their close binary nature has required significant observational effort \citep{Brown2022}. Therefore, it is unsurprising that there is a qualitative change in both WD-MS and WD-WD binary fractions in Fig.~\ref{fig:wbf-mass} at $m=0.6$\Msun: most of the white dwarfs that contribute to these curves at $<0.6$\Msun\ must be in hierarchical triple systems, whereas most of the white dwarfs above these mass are likely not. 

\begin{figure}
	\centering
	\includegraphics[width=1\linewidth]{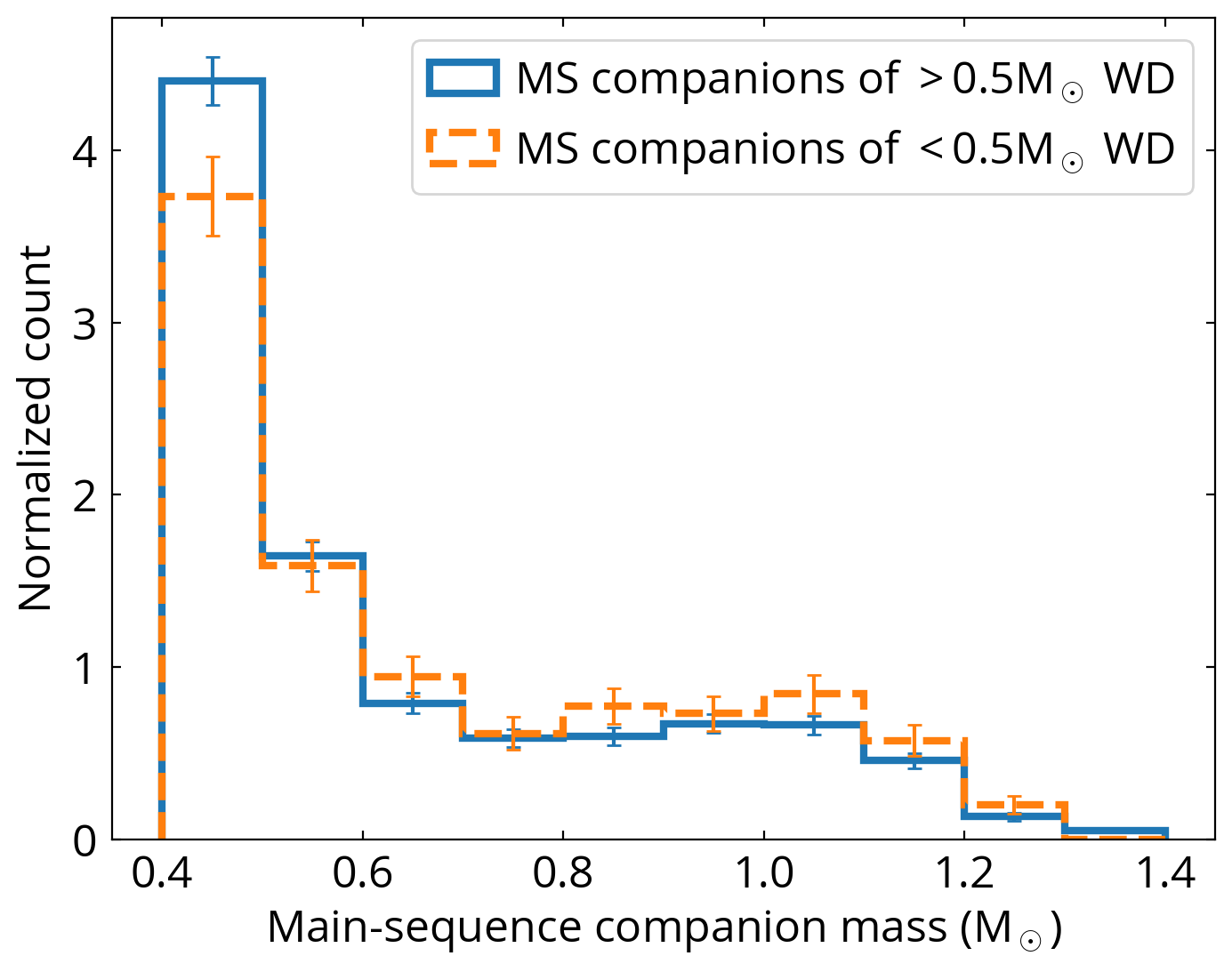}
	\caption{The mass distribution of the main-sequence companions to the white dwarfs in wide WD-MS binaries split into two bins of mass. Low-mass white dwarfs have wide companions that have slightly higher mass than high-mass white dwarfs. We have also checked different WD mass bins above 0.5\Msun\ and did not find any differences in the mass distributions of their MS companions. A conservative parallax$>4$ mas cut was used to mitigate any possible selection effects. The error bars are statistical based on the number of objects contributing to each bin.}
	\label{fig:MSmass}
\end{figure}

Interestingly, Fig.~\ref{fig:wbf-mass} suggests that the wide companion of a low-mass white dwarf is much more likely to be a main-sequence star rather than a white dwarf, compared to a single high-mass white dwarf. Furthermore, Fig.~\ref{fig:MSmass} shows that the MS mass of low-mass ($m_{\rm WD}<0.5$\,\Msun) white dwarfs is more massive than that of the higher-mass ($m_{\rm WD}>0.5$\,\Msun) white dwarfs. The higher mass of the wide MS companions is not due to the younger stellar ages, because the kinematics (e.g. space velocities) suggest that low-mass WDs are actually kinematically older than the high-mass WDs. 

Although we do not include hierarchical triples in our model in Sections \ref{sec:model} and \ref{sec:results}, here we discuss some interesting possible effects qualitatively. Close MS-MS binaries with orbital periods $<10$\,days have an enhanced wide MS companion fraction out to $10^3$-$10^4$\,AU \citep{Hwang2023}. If the inner close binary later evolves to a WD-WD binary, its light will be dominated by the lower-mass, larger luminosity WD and in observations it will appear as a low-mass WD with a wide companion at $\sim10^3$\,AU, resembling the sample in this analysis. The higher MS mass of companions to such low-mass WDs may be due to the Kozai-Lidov effect \citep{Kozai1962, Lidov1962}, where a more massive tertiary has a larger dynamical effect on the inner binary. However, when the tertiary is too massive, its MS lifetime is too short and the inner binary does not have a chance to finish its Kozai-Lidov dynamical process during this time, which may contribute to the low WD-WD binary fraction for low-mass white dwarfs. Another possible effect is that a more massive tertiary may make it more likely that the triple survives the mass-loss episode associated with the common envelope evolution. Future investigations taking into account all these interesting contributions are needed to fully understand the mass dependence of the white dwarf binary fraction in the low-mass regime. 

\subsection{Retention fraction}

\begin{figure}
	\centering
	\includegraphics[width=1\linewidth]{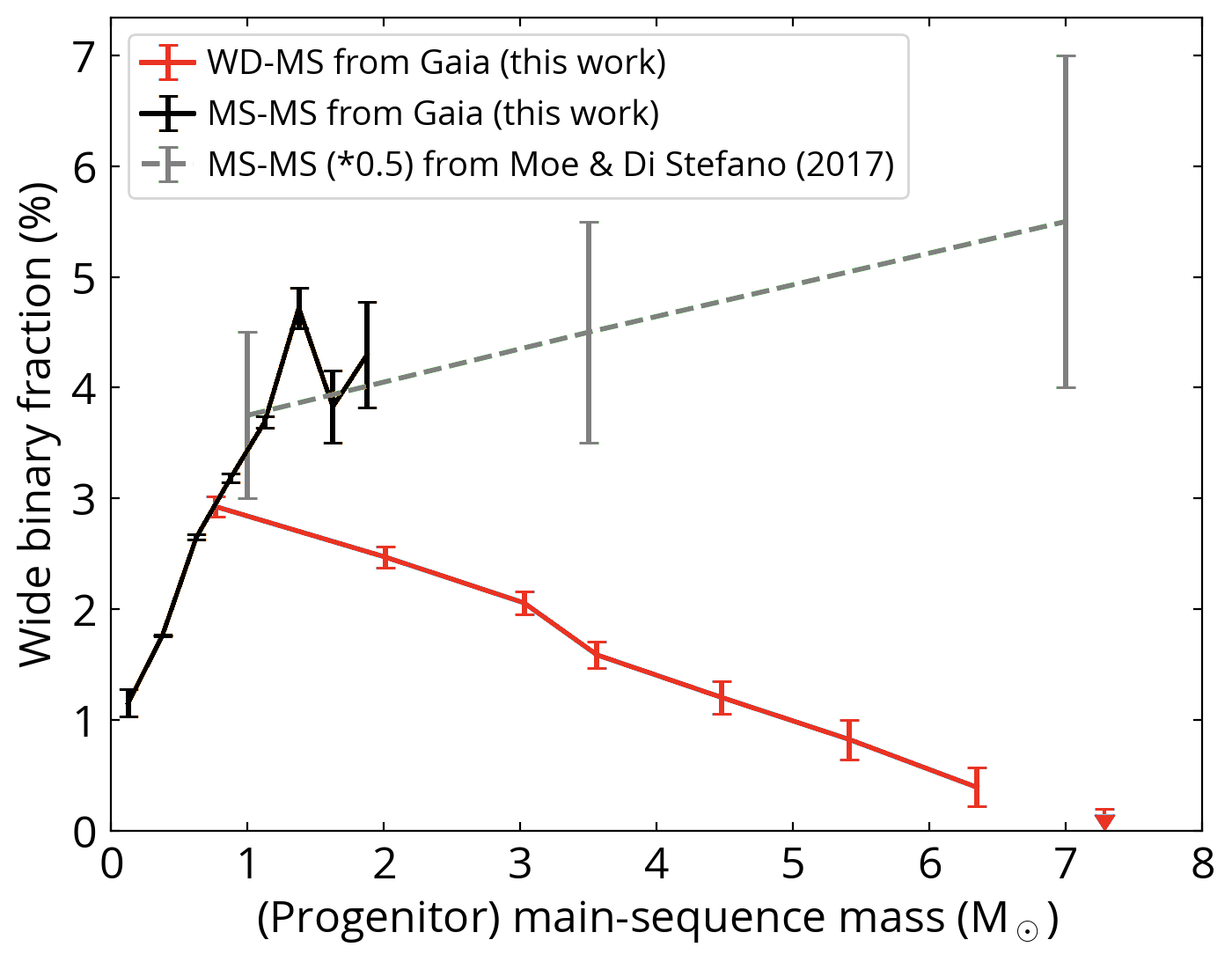}
	\caption{MS-MS (black and grey) and WD-MS (red) mass-dependent wide binary fractions. The WD-MS binary fraction is shown as a function of the mass of the progenitor to the observed WD as estimated using the \citet{Cummings2018} initial-to-final mass relationship. The low-mass MS-MS binary fraction is from our {\it Gaia}-based measurements, and the high-mass MS-MS binary fraction is from \citet{Moe2017} normalized to agree with the {\it Gaia} value at 1\Msun.}
	\label{fig:WBF-MS1}
\end{figure}

\begin{figure}
	\centering
	\includegraphics[width=1\linewidth]{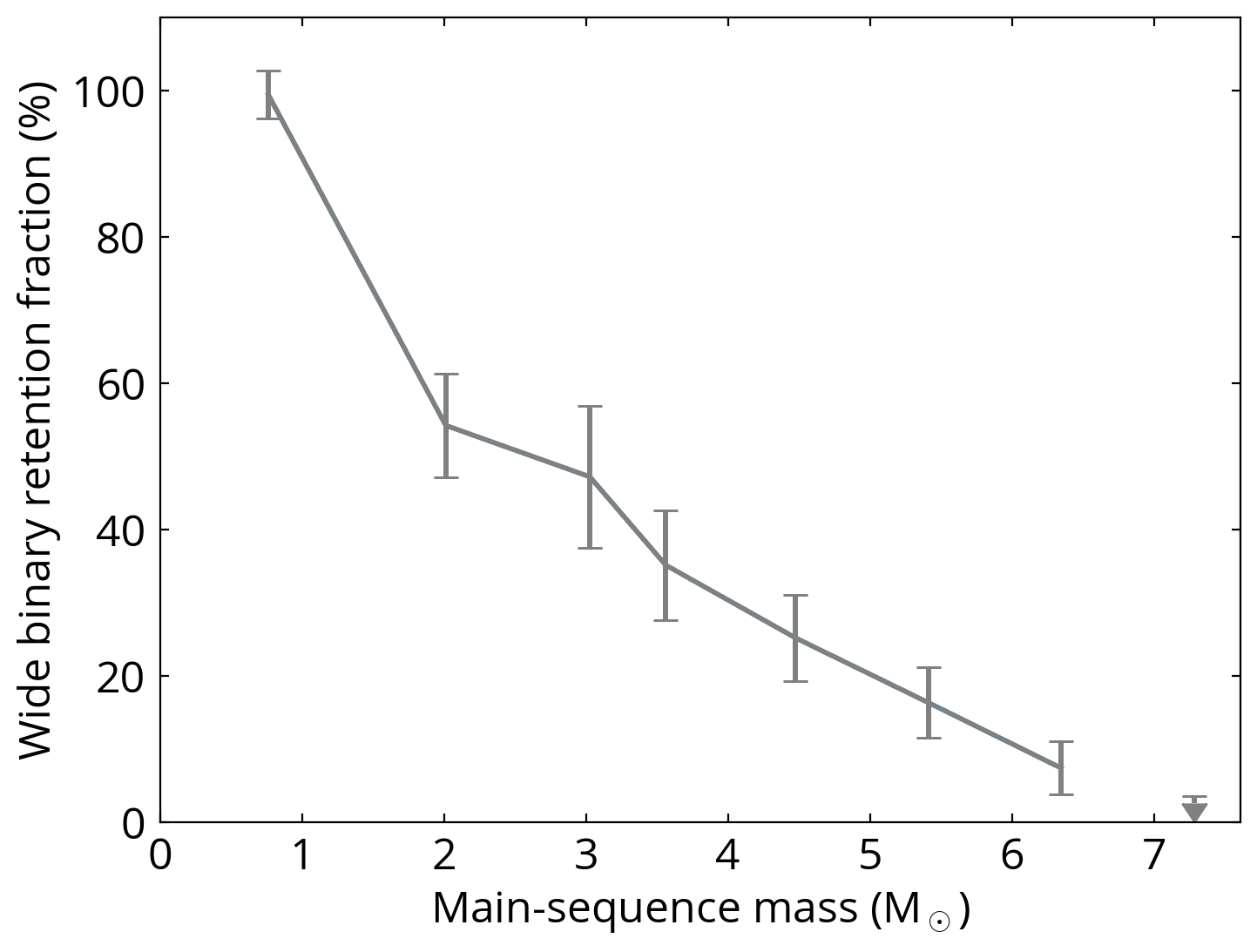}
	\caption{The empirical retention fraction of wide binaries when a star with different masses becomes a white dwarf.}
	\label{fig:retention}
\end{figure}

In this section, we measure the empirical retention fraction -- the probability for a star to keep its wide companion when it evolves from a main-sequence star to a white dwarf. We expect that the retention fraction provides constraints on the stellar evolution, including the mass loss process and the velocity recoil. Figures \ref{fig:WBF-MS1} and \ref{fig:retention} show the steps in this measurement. For every bin in white dwarf mass, we have measured the wide binary fraction $f_{\rm WD-MS}$ in this bin (Fig.~\ref{fig:wbf-mass}). Then we estimate the mass of the main-sequence progenitors for this WD mass bin using the initial-to-final mass relationship from \citet{cumm08}, and for the corresponding main-sequence mass we can measure the MS-MS wide binary fraction $f_{\rm MS-MS}$, either from {\it Gaia} data or from the literature (Fig.~\ref{fig:WBF-MS1}). The empirical retention fraction $R_{\rm ret}$ is then defined as the ratio of the two, $f_{\rm WD-MS}/f_{\rm MS-MS}$. 

The wide binary fractions are all defined in the same separation range $10^3-10^4$ AU. If the only possible effect of the mass loss and velocity recoil in an MS-MS binary was orbital disruption, without any changes to the semi-major axis for the surviving binaries, then the empirical retention fraction would be just the same as the actual retention fraction during the mass-loss process. In practice, mass loss and recoil result in the evolution of all orbital parameters of surviving binaries, so the WD-MS binaries have MS-MS progenitors with a different separation distribution. These effects are all properly taken into account in Sections \ref{sec:model} and \ref{sec:results} when we compare the theoretical retention fractions with the empirical ones.  

To complete the calculation of the retention fraction, we now need the wide binary fraction of the MS-MS binaries. Fig.~\ref{fig:WBF-MS1} shows the MS-MS wide-binary fraction from {\it Gaia}. The advantage of using {\it Gaia} over any previous measurements of this binary fraction is that our MS-MS and WD-MS samples have the same comoving selections and share similar systematics. For all wide binary fractions in Fig.~\ref{fig:wbf-mass} and Fig.~\ref{fig:WBF-MS1}, we limit the selection of the MS companion to $G<10.5$ mag and BP-RP$>0.8$ mag. Using this relatively red selection reduces the effects of stellar evolution on the measured wide-binary fractions, but none of our results are qualitatively different when these cuts are removed or modified. 

However, {\it Gaia} has very few high-mass MS stars in our 500-pc sample. Therefore, in Fig.~\ref{fig:WBF-MS1}, we supplement our measurements with those of the wide binary fraction from \citet{Moe2017}. Because our sample from {\it Gaia} and the sample from \citet{Moe2017} have very different companion completeness, the fractions from \citet{Moe2017} are manually multiplied by a factor such that their fractions agree with our measurements from {\it Gaia} at 1$\rm M_\odot$. In other words, we use the absolute values of the wide-binary fraction from {\it Gaia} defined according to eq. (\ref{eq:bfraction}), and we use the overall mass trend from \citet{Moe2017} to extend the wide-binary fraction values to the high-mass end. Because the \citet{Moe2017} measurements are given as a function of period and not the separation, there could be some subtle mass-dependent effects in matching these values to ours, but the number of objects is so low and the error bars are so high that we assume these types of uncertainties are subdominant to the small number statistics. It does however seem that from the combination of {\it Gaia} and \citet{Moe2017} measurements that the MS-MS wide-binary fraction is a steeply rising function of mass. 

Fig.~\ref{fig:retention} shows the retention fraction computed by diving the WD-MS wide binary fraction (red) by the MS-MS wide binary fraction (black+grey) in Fig.~\ref{fig:WBF-MS1}. The error bars reflect the uncertainties of the WD-MS wide-binary fraction. The overall trend is reliable, but the exact values may be subject to the systematic uncertainties in the WD-MS wide-binary fraction -- for example, what kinds of companions are and are not included -- and thus could be different by some normalization factor with a different sample selection.  

\subsection{Eccentricities}

\begin{figure*}
	\centering
	\includegraphics[width=0.45\linewidth]{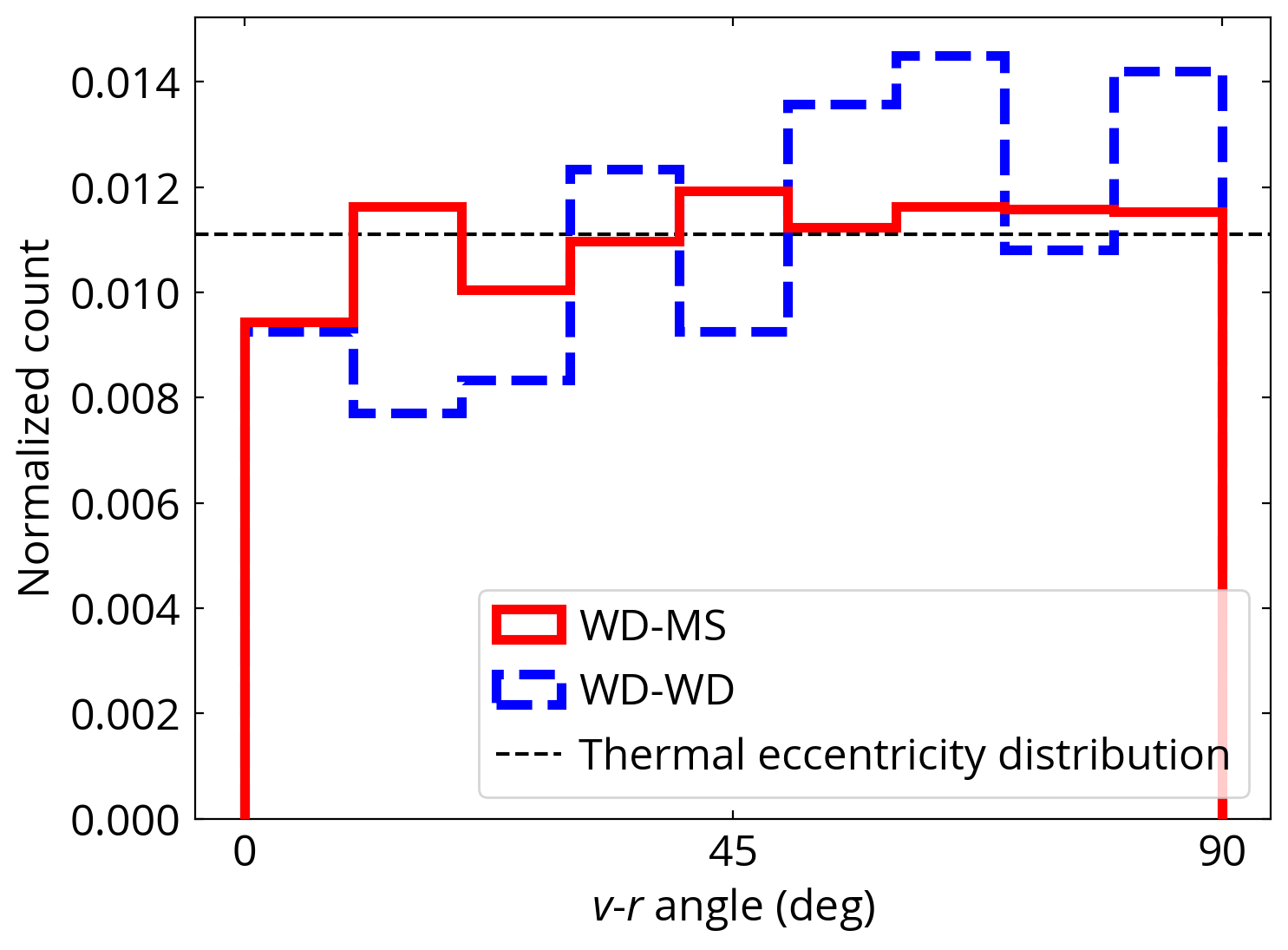}
	\includegraphics[width=0.45\linewidth]{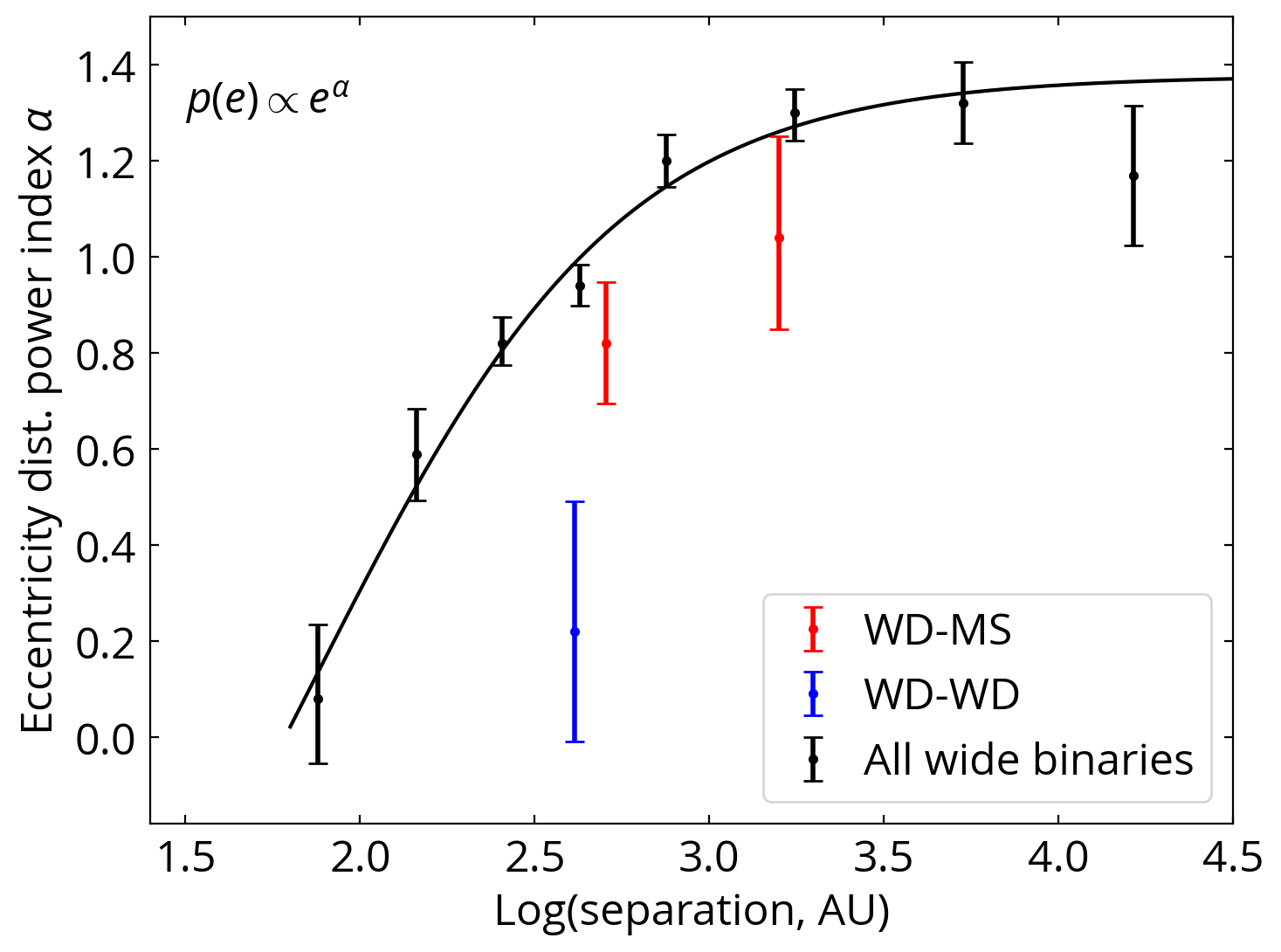}
	\caption{Eccentricity measurements for WD-MS and WD-WD wide binaries. Left: raw $v-r$ angle distributions for a model population with a thermal eccentricity distribution ($p(e)=2e$, horizontal dashed black line), and for WD-MS and WD-WD wide binaries (red and blue, respectively). Right: the best-fit power-law indices of the eccentricity distribution $p(e)\propto e^{\alpha}$ for all wide binaries (black data points with error bars and functional fit from \citealt{Hwang2022ecc}) and for WD-MS (red) and WD-WD (blue) binaries. }
	\label{fig:alpha_sep}
\end{figure*}

In addition to high-quality distances and photometry, {\it Gaia} also yields proper motions. The wide binaries in this paper have been selected by requiring that the two stars must have similar proper motions in order to be considered co-moving, but these proper motions are not identical because of the relative orbital velocity between them projected on the plane of the sky. The angle between the relative velocity vector and the vector connecting the two stars -- called the $v-r$ angle -- carries information about the eccentricity of the binary \citep{Tokovinin1998}: for example, for a circular orbit in the plane of the sky this value is always $90$ deg. While individual eccentricities are impossible to measure from one snapshot due to the unknown inclination of the orbit, it is possible to measure the eccentricity distribution of a large population from its distribution of the $v-r$ angles (Fig.~\ref{fig:alpha_sep}, left; \citealt{Hwang2022ecc}).

Using this technique, \citet{Hwang2022ecc} demonstrated that the typical eccentricities of MS-MS binaries are a strong function of separation. A power-law distribution function $p(e)\propto e^{\alpha}$ turns out to be a good description of the data across a broad range of separations, with MS-MS binaries showing uniform eccentricity distribution at $<100$ AU and increasing values of $\alpha$ at higher separations (Fig.~\ref{fig:alpha_sep}, right). Other than `twin' wide binaries which have eccentricities much higher than other wide binaries at similar separations, there is no significant dependence of eccentricities on mass ratios \citep{Hwang2022twin}.

We use the $v-r$ angle technique to measure the eccentricity distributions of WD-MS and WD-WD wide binaries. Specifically, we use the classification of WD-MS and WD-WD from the Gaia wide binary catalog \citep{El-Badry2021}, and we further require that parallaxes $>5$\,mas, the significance of the proper motion difference $>3$, probability of chance-alignment pair $<0.1$, and angular separations $>1.5$\,arcsec to avoid Gaia systematics \citep{Hwang2022ecc}. No additional photometry criteria are imposed here. For WD-MS binaries, we use two bins of separations, [100, 1000] and [1000, 3000]\, AU, with the median separations of $10^{2.70}$ and $10^{3.20}$\,AU in each bin. For WD-WD binaries, we only consider one separation bin between 100 and 1000\,AU due to its smaller sample size, with a median separation of $10^{2.61}$\,AU. At larger separations, the orbital velocities are small and the size of the sample that has sufficient non-zero proper motion difference is too limited for the eccentricity analysis.  

The left panel of Fig.~\ref{fig:alpha_sep} shows the raw distributions of $v-r$ angles for WD-MS (red) and WD-WD (blue) binaries at $10^2-10^3$\,AU separations. By definition, $v-r$ angles range between 0 and 180\,deg, but here we fold it at 90\,deg (e.g., 100\,deg would be folded to 80\,deg) since a Keplerian orbit has a symmetric $v-r$ angle distribution with respect to 90\,deg. Compared to the theoretical $v-r$ angle distribution (dashed black horizontal line) from the thermal eccentricity distribution, WD-WD binaries (blue) have some deficit of $v-r$ angles close to 0 (WD-MS binaries do as well, but to a lesser degree), suggesting a slightly sub-thermal eccentricity distribution. The right panel of Fig.~\ref{fig:alpha_sep} shows that WD-MS binaries have $\alpha$ values that are significantly lower than those of MS-MS binaries at the same separation, and WD-WD binaries have $\alpha$ lower still. The error bars are 68\% credible intervals of $\alpha$, and the horizontal positions are the median separations of the sample. 

Along with the steeply declining wide binary fraction as a function of white dwarf mass, the relatively low eccentricities of the WD-MS and WD-WD binaries are another observational puzzle we would like to be able to explain.

\section{Orbital evolution due to mass loss}
\label{sec:model}

Our goal is to develop a model that connects stellar mass loss processes with the dynamical evolution of the wide binary population. We consider a binary of two stars, one with mass $m_1$ and the other with mass $m_2$. Star $m_1$ is assumed to be non-varying, whereas star $m_2$ may suffer mass loss on a variety of timescales -- shorter than the orbital period or longer than the orbital period -- as well as a velocity kick if the mass loss is not spherically symmetric. In this section we discuss how to compute the resulting changes in the orbital semi-major axis and eccentricity. 

\subsection{Mass loss processes and their timescales}

For a 1\Msun$-$1\Msun\ binary, the orbital period at $10^3$ AU is $2\times10^4$ years, and at $10^4$ AU it is $7\times 10^5$ years. At the $10-40$\kms\ expansion rate of the post-asymptotic-giant-branch (post-AGB) stellar wind \citep{gold17, hofn18}, the stellar wind produced during this stage of stellar evolution is expected to leave the orbit in $1000-5000$ years (for $10^3$ AU) or in $10-50$ kyr (for $10^4$ AU), significantly faster than the orbital timescale. If the mass is lost in a ``common envelope" event, then the expansion velocities are higher, $100-1000$ \kms\ \citep{ivan13a}. Therefore, for wide binaries we can assume that the mass, once ejected, leaves the orbit instantaneously and does not exert gravitational forces on the stars. 

In contrast, the characteristic timescale over which mass $m_2$ changes appreciably varies widely. The timescale for the mass change in a common envelope event is negligible in comparison to the orbital timescales \citep{ivan13b}. The time dependence of the (post-)AGB mass loss rate is not well known, but models and observations generally suggest that it is an accelerating function of time: mass loss proceeds relatively slowly for tens to hundreds of thousands of years and ends with an episode of relatively rapid mass loss. For example, in models by \citet{stef98}, a 2.8\Msun\ star loses 50\% its mass in $3\times 10^5$ years and then the remnant loses another 50\% of its mass in $10^4$ years on its way to becoming a white dwarf. In another example by \citet{weis09}, a 1.9\Msun\ star loses 30\% over 2 Myr, and then the remnant loses 70\% over $7\times 10^4$ years in a rapidly accelerating fashion. Both the AGB phase and the post-AGB phase may be important for the net mass loss, and the result is a steep dependence of the timescale on mass. AGB models by \citet{weis09} demonstrate that stars with initial mass $<2$\Msun\ finish their AGB phase with masses nearly equal to their core masses, so their timescale for mass loss is the AGB timescale ($10^5-10^6$ years). In contrast, more massive stars finish their AGB phase still with tens of per cent of their mass to lose on the post-AGB timescales, which themselves are a very steeply declining function of mass, from $10^5$ years to under $10^3$ years for the initial masses $0.8-4$\Msun\ \citep{mill16}.

Mass loss which is not exactly centrally symmetric imparts a net kick recoil velocity to the remnant. The kick for white dwarf remnants of single stellar evolution has been estimated to be as high as 4 \kms\ \citep{freg09,izza10} from indirect measurements of white dwarf distributions in globular clusters and relatively close binary systems, but recently was revised down to $\sim 0.75$ \kms\ and constrained to be $<2$ \kms\ based on the wide binary separation distribution \citep{elba18}. Given that orbital velocities are $\lesssim 1$ \kms\ at $>10^3$ AU, kicks of this magnitude might be important for the orbital configuration and disruption. We revisit these constraints in this paper in the context of the observations presented above and with a more detailed model for the velocity recoil. 

The relative motion of an $m_1+m_2$ binary with separation $\vec{r}(t)$ (vector directed from $m_2$ to $m_1$) subject to gravitational interaction is described by $\ddot{\vec{r}}=-G(m_1+m_2)\vec{e}_r/r^2$, where $\vec{e}_r$ is the unit vector along $\vec{r}$. If $m_2$ is losing mass and recoiling -- or receiving velocity kicks $\vec{v}_k$ -- due to some internal processes not coupled to the orbit (e.g., asymmetric wind), the center of mass of the binary becomes a non-inertial frame, but taking this into account one can still formulate the equation for the evolution of the separation vector,
\begin{equation}
\ddot{\vec{r}}=-\frac{G(m_1+m_2(t))}{r^2}\vec{e}_r-\frac{{\rm d}\vec{v}_k}{{\rm d} t}.
\label{eq:motion}
\end{equation}

In reality, $\vec{v}_k$ could be randomly varying in direction if the processes that launch the wind (stellar pulsations and radiative pressure) have stochastic variations that are not centrally-symmetric. Here we only consider models where the direction of $\vec{v}_k$ remains the same throughout the evolution of $m_2$. This may be the case if for example the stellar wind has a net momentum along some preferred axis, most likely the stellar axis of rotation. We qualitatively expect that at fixed observational constraints on net $|\vec{v}_k|$, the models with stochastic wind directions can accommodate larger individual kicks (since they can partially cancel each other out if they are directed opposite one another) than models with uni-directional recoil. 

Although it is commonly called a kick velocity, for a single star evolving into a white dwarf the recoil develops over the same timescales as those over which the mass loss proceeds \citep{izza10}. Therefore, we tie the velocity kick to the fractional mass loss using momentum conservation, 
\begin{equation}
{\rm d}v_k/{\rm d} t =v_{\rm asym} \dot{m}_2(t)/m_2(t).   
\label{eq:vkick}
\end{equation}
Here $v_{\rm asym}$ quantifies the degree of asymmetry of the mass loss and is a major adjustable parameter of our model. We envision that $v_{\rm asym}=\epsilon v_{\rm wind}$, where $v_{\rm wind}\sim 10-40$ km/sec is the net wind velocity and $\epsilon$ is its measure of asymmetry. If the wind is centrally-symmetric, $\epsilon=0$ and there is no recoil. The parameter $v_{\rm asym}$ is related to the net kick velocity constrained for example by observations of globular clusters \citep{freg09,izza10} by the total fractional mass loss, $v_k=v_{\rm asym}\ln (m_{2, {\rm initial}}/m_{2, {\rm final}})$.  

\subsection{Population synthesis model setup}
\label{sec:pop}

We set up a binary population synthesis model similar to that of \citet{elba18}, with some interesting differences informed by recent observations. \citet{Moe2017} demonstrated that for wide binaries, the masses of the components are largely uncorrelated. Therefore, similarly to \citet{elba18}, we draw stellar masses independently from the \citet{krou01} initial mass function (with minimal mass 0.3\Msun\ and maximal mass 7.2\Msun) and periods from \citet{duqu91} and \citet{fisc92}. We uniformly draw the binary's birthday from the last 12 Gyr (equivalent to setting a constant star-formation rate in this period) and for each binary determine whether zero, one or both of the stars have evolved into a white dwarf by the present day based on the masses of stars and their corresponding main-sequence lifetimes. We use main-sequence lifetime as a function of mass from \citet{lame17}. 

Wide binaries with white dwarfs show a puzzling eccentricity distribution which is quite different from that of main-sequence binaries, and therefore we are particularly interested in a realistic setup and evolutionary model for eccentricities. In contrast to \citet{elba18}, we do not use the uniform eccentricity distribution at all semi-major axes. Instead, at each semi-major axis we draw from the power-law eccentricity distribution $p(e)=e^{\alpha}/(1+\alpha)$, where $\alpha$ as a function of separation is provided by \citet{Hwang2022ecc}, and we ignore the differences between intrinsic semi-major axes and observed separations for the purposes of this calculation. Furthermore, in contrast to \citet{elba18} we do not attempt to properly follow the close binary evolution on the assumption that the binaries that start at semi-major axes $\ll 100$ AU rarely contribute to the population that ends up at $10^3-10^{4.5}$ AU we are focused on here. Finally, in the model of \citet{elba18}, the kick occurs instantaneously at the end of the mass loss and is drawn from a distribution which is independent of the amount of mass lost. In our model, the recoil is directly tied to the mass loss during the evolution via eq. (\ref{eq:vkick}). 

For each binary we line up the $z$ axis with the direction of $\vec{v}_k$ and randomly draw the remaining orbital elements -- inclination $i$, argument of the pericenter $\omega$, longitude of the ascending node $\Omega$ and mean anomaly $M$ -- from the appropriate distributions \citep{murr99} and consider this setup to be the initial conditions for the phase of the (post-)AGB mass loss when eq. (\ref{eq:motion}) describes the orbital evolution. 

Finally, we also set up a comparison population of singles, assuming a uniform star formation rate over the last 12 Gyr and \citet{krou01} initial mass function. At present day, the singles are either main-sequence stars or white dwarfs depending on whether their mass-dependent main-sequence phase is longer or shorter than their lifetime to date. The number ratio $N_{\rm singles}/N_{\rm binaries}$ (corresponding to the fraction of stars born in binaries vs singles) is a major adjustable parameter of our model. When evaluating the results of the population model, we add to the population of singles any disrupted wide binaries, as well as close binaries with $a<100$ AU -- for these objects, we assume that they appear to the observer as a single star with the luminosity of the brighter component. 

\citet{Hwang2022ecc} found that the distribution of wide binary eccentricities is well-fit by $p(e|\alpha)=(1+\alpha)e^{\alpha}$, with the parameter $\alpha$ varying as a function of binary separation. As we parameterize the observed population using this distribution function and the corresponding $\alpha$, we have a need to measure $\alpha$ in our population synthesis model. To this end, for any set of eccentricities $e_i$ we can formally maximize the likelihood 
\begin{equation}
  \mathcal{L}=\prod_{i=1}^N p(e_i|\alpha) =(1+\alpha)^N \left(\prod_{i=1}^N e_i\right)^{\alpha}   
\end{equation}
by solving the equation $\partial \ln \mathcal{L}/\partial \alpha=0$ to find
\begin{equation}
    \alpha=-\frac{N}{\sum \ln e_i}-1.
\end{equation}
In practice we find that mass loss and kicks preferentially disrupt high-eccentricity orbits, so a power-law distribution over the entire range of bound eccentricities $e\in [0,1)$ is often a poor fit to the post-evolution populations. To capture both the overall trend and the systematic uncertainty, we compute $\alpha$ over the range of eccentricities from 0 to $e_{\rm max}\le 1$ with several different values of $e_{\max}$ (typically 0.7, 0.8, 0.9 and 1.0) and report the median $\alpha$ computed from these measurements as well as the range as an error bar in $\alpha$. 

\subsection{Computing the change of orbit due to mass loss and kicks}

With the binaries and singles set up as described above, we compute orbital evolution described by eq. (\ref{eq:motion}). The total amount of mass loss at each main-sequence starting mass is given by the initial-to-final mass relationship of \citet{Cummings2018}. A variety of functions can be used to describe the mass loss rate as a function of time, and this dependence is quite poorly known {\it a priori}. If we assume that mass loss proceeds at a constant rate, then $m_2(t)$ linearly declines, whereas $v_k(t)$ is then given by eq. (\ref{eq:vkick}). The only parameter necessary to describe this evolution is then $\tau_{\rm AGB}$, and we explore several options for this parameter. 

Orbital evolution in response to mass loss and velocity kicks is qualitatively different depending on the relationships between three timescales of the problem: the orbital period $\tau_{\rm orb}$, the evolutionary timescale $\tau_{\rm AGB}$ and the secular timescale $\tau_{\rm sec}$. Evolution is secular when $\tau_{\rm AGB}\gg \tau_{\rm orb}$ and therefore the extra force due to the recoil can be orbit-averaged, and the effective Hamiltonian for the perturbation due to the recoil is $H_1=\dot{\vec{v}}_k\cdot \langle\vec{r}\rangle$. The absolute value of $\langle \vec{r}\rangle$ is $3ea/2$, where $e$ is the orbital eccentricity and $a$ is the semi-major axis, and it is directed to the apocenter. 

In the secular regime our problem is similar to the classical Stark problem \citep{heis86, bely10}, where a gravitationally bound particle is subject to an additional weak constant force, except we additionally allow for changes in mass and force magnitude. The problem can be reduced to two coupled ordinary differential equations in Delaunay variables \citep{heis86, bely10}: 
\begin{eqnarray}
\frac{{\rm d}\omega}{{\rm d} t}=\frac{\partial H_1}{\partial J};\label{eq:hamiltonian1}\\
\frac{{\rm d}J}{{\rm d} t}=-\frac{\partial H_1}{\partial \omega},\label{eq:hamiltonian2}
\end{eqnarray}
with the right hand sides explicitly calculable from the Hamiltonian in Delaunay variables:
\begin{equation}
    H_1=-\dot{v}_k\frac{3L^2}{2Gm}\sqrt{1-\frac{J^2}{L^2}}\sqrt{1-\frac{J_z^2}{J^2}}\sin\omega.
\end{equation}
Here $\omega$ is the argument of the pericenter, $m(t)=m_1+m_2$ is the total mass of the system, $L=\sqrt{G(m_1+m_2)a}$ is an integral of motion, $J=L\sqrt{1-e^2}$ is the angular momentum per reduced mass and $J_z=J\cos i$ is its projection on the $z$ axis fixed by the direction of $\vec{v}_k$ and is also an integral of motion. 

In the classical Stark case of constant mass and constant acceleration in eq. (\ref{eq:motion}), the system described by eq. (\ref{eq:hamiltonian1})-(\ref{eq:hamiltonian2}) moves on closed loops in the $e-\omega$ space on timescale $\tau_{\rm sec}\sim \frac{1}{\dot{v}_k}\sqrt{G(m_1+m_2)/a}$. The tightest binaries have $\tau_{\rm sec}\gg\tau_{\rm AGB}$ and they barely have time to travel around their loops during the (post-)AGB evolution and therefore their eccentricities change very little. The only effect that the mass loss and the associated recoil have on these binaries is the adiabatic orbital expansion, $a(t)\propto 1/(m_1+m_2(t))$, as required by the conservation of Delaunay action $L$. But for binaries of interest to our problem -- the ones that end up with separations in the $10^3-10^{4.5}$ AU range -- the secular evolution equations (\ref{eq:hamiltonian1})-(\ref{eq:hamiltonian2}) need to be integrated numerically to compute the change in eccentricities. In practice, we numerically integrate these equations for all binaries with $\tau_{\rm AGB}>30 \tau_{\rm orb}$ regardless of their $\tau_{\rm sec}$. In our typical population synthesis model, $\sim$50\% of the WD-MS binaries are in this regime, and under the assumption that the mass loss is slow enough such binaries cannot get disrupted.

In the opposite case, $\tau_{\rm AGB}\ll \tau_{\rm orbital}$, the mass loss and the kick are essentially instantaneous compared to the orbital timescale. In this ``impulse" limit, the change of the orbit can be calculated analytically for any values of the orbital parameters at the time of the kick. The orbit can be disrupted due to mass loss alone, even if there is no kick. For the binaries wide enough that the condition $\tau_{\rm AGB}\ll \tau_{\rm orbital}$ is satisfied, the kick velocities of a few km/sec are quite significant compared to the orbital velocities. In practice, we use the impulse approximation for $\tau_{\rm AGB}<0.1 \tau_{\rm orbital}$, and $>95\%$ of these binaries are disrupted, but in a typical population synthesis model only $\sim 10$\% of the WD-MS binaries are in this regime. 

For the remaining $\sim 40$\% of the WD-MS binaries, $\tau_{\rm AGB}$ is in the range $0.1-30\tau_{\rm orb}$, and direct numerical integration of the equation of motion (\ref{eq:motion}) is required (this type of separation of computation methods based on timescales was also employed e.g. by \citealt{Pham2024} for another application relevant to white dwarfs). Because the calculations only need to cover a few orbits (otherwise the binary would be in the secular regime), we do not have concerns about long-term integration accuracy. We prioritize speed and use a second order leapfrog integrator \citep{binn08} with at least 100 steps per orbit or per $\tau_{\rm AGB}$, whichever is the finer temporal resolution. REBOUND \citep{rein12} is a thoroughly tested orbital integrator which can handle external acceleration, but not changing mass. Therefore, we use REBOUND to test our leap-frog integrator on a classical Stark problem and verify that we obtain accurate results for the final eccentricities and accurately predict disruption. Tens of per cent of binaries in this period range are disrupted, depending on the value of $v_{\rm asym}$. 

\section{Simulation results and trends}
\label{sec:results}

Our model has a large number of adjustable parameters. The three key values are $v_{\rm asym}$, $\tau_{\rm AGB}$ and $N_{\rm binaries}/N_{\rm singles}$. However, the initial distribution of periods or semi-major axes, the star formation history, and the initial eccentricity distributions can also be easily varied. To probe these parameters, we have a range of interesting observables: the distribution of semi-major axes for MS-MS, WD-MS and WD-WD binaries (from \citealt{elba18}), eccentricity distributions (Fig.~\ref{fig:alpha_sep}), binary fractions and the related retention fractions as a function of mass (Figures \ref{fig:WBF-MS1} and \ref{fig:retention}), and overall relative numbers of MS-MS, WD-MS and WD-WD binaries. In principle, for the correct model all of these observables should be reproduced simultaneously. In practice, some of the parameters of the model are known much better than others, and furthermore not all parameters of the model have a strong impact on the predicted observables. Therefore, instead of looking for the best-fit model in any statistical sense, in this section we describe the range of parameters we have explored and their effects and the choices we made for our most successful models. 

\subsection{Main sequence - main sequence (MS-MS) binaries}

We first consider the binary fraction of main sequence stars. For this purpose, only the population synthesis setup described in Section \ref{sec:pop} is relevant and the subsequent orbital calculations are unnecessary: while the stars are on the main sequence, their mass loss is negligible and therefore we assume that their orbital periods and eccentricities remain the same as at birth. The only reason that the present-day distributions of orbital parameters may be different from those at birth is that some stars have evolved away from the main sequence and therefore no longer contribute either to the MS-MS binary population or to the MS singles. 

\begin{figure*}
    \centering
    \includegraphics[width=5.8cm]{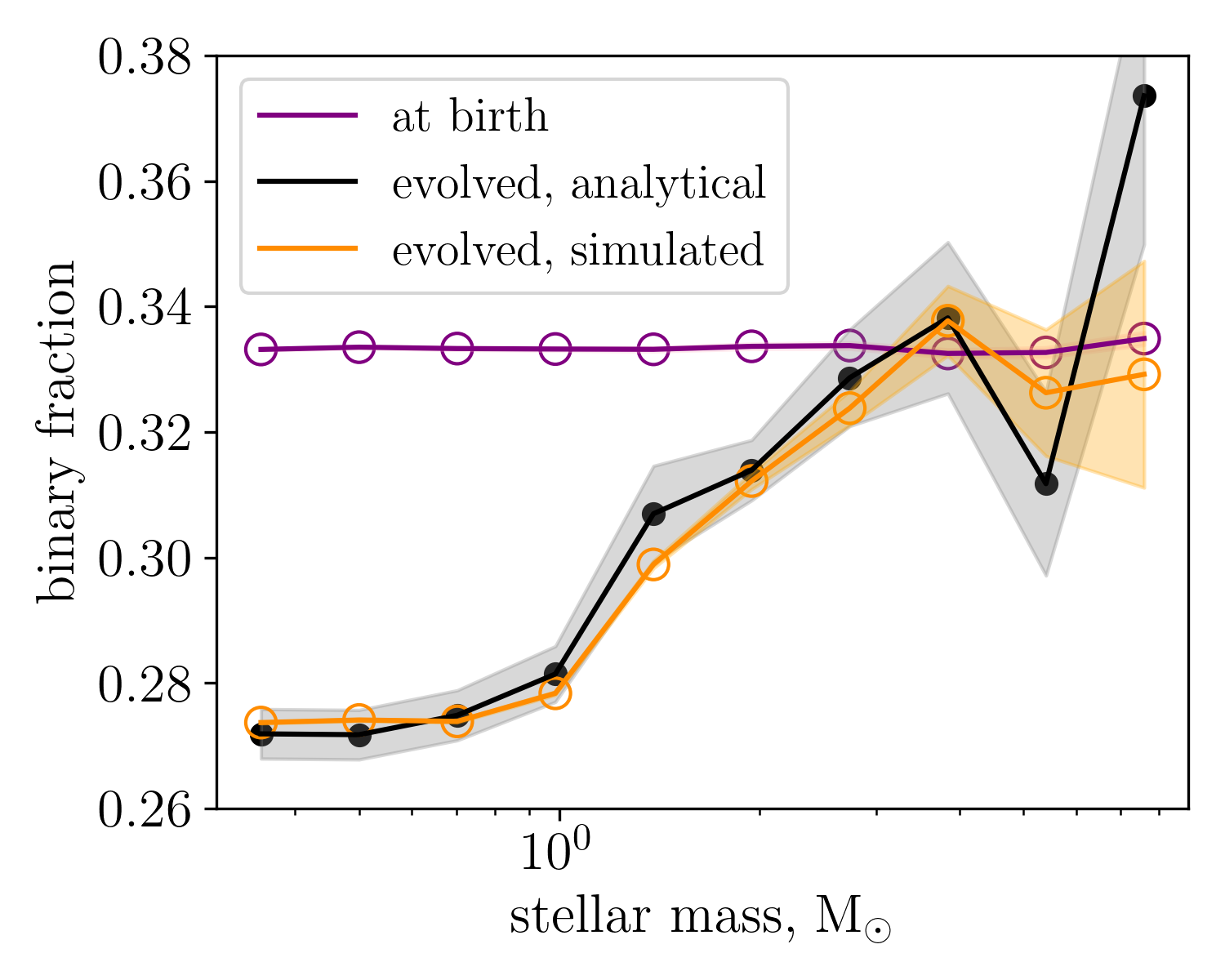}
    \includegraphics[width=5.8cm]{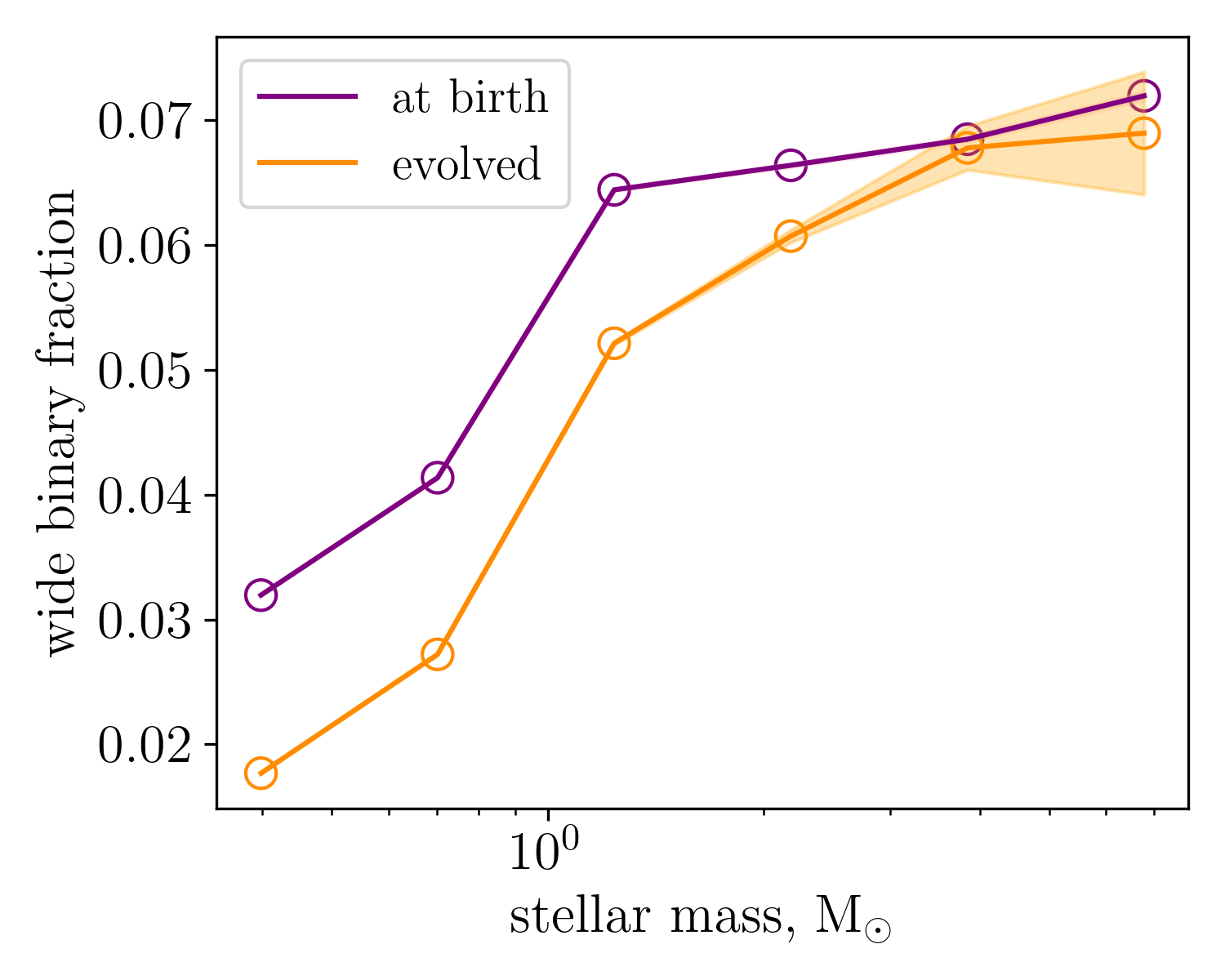}
    \includegraphics[width=5.8cm]{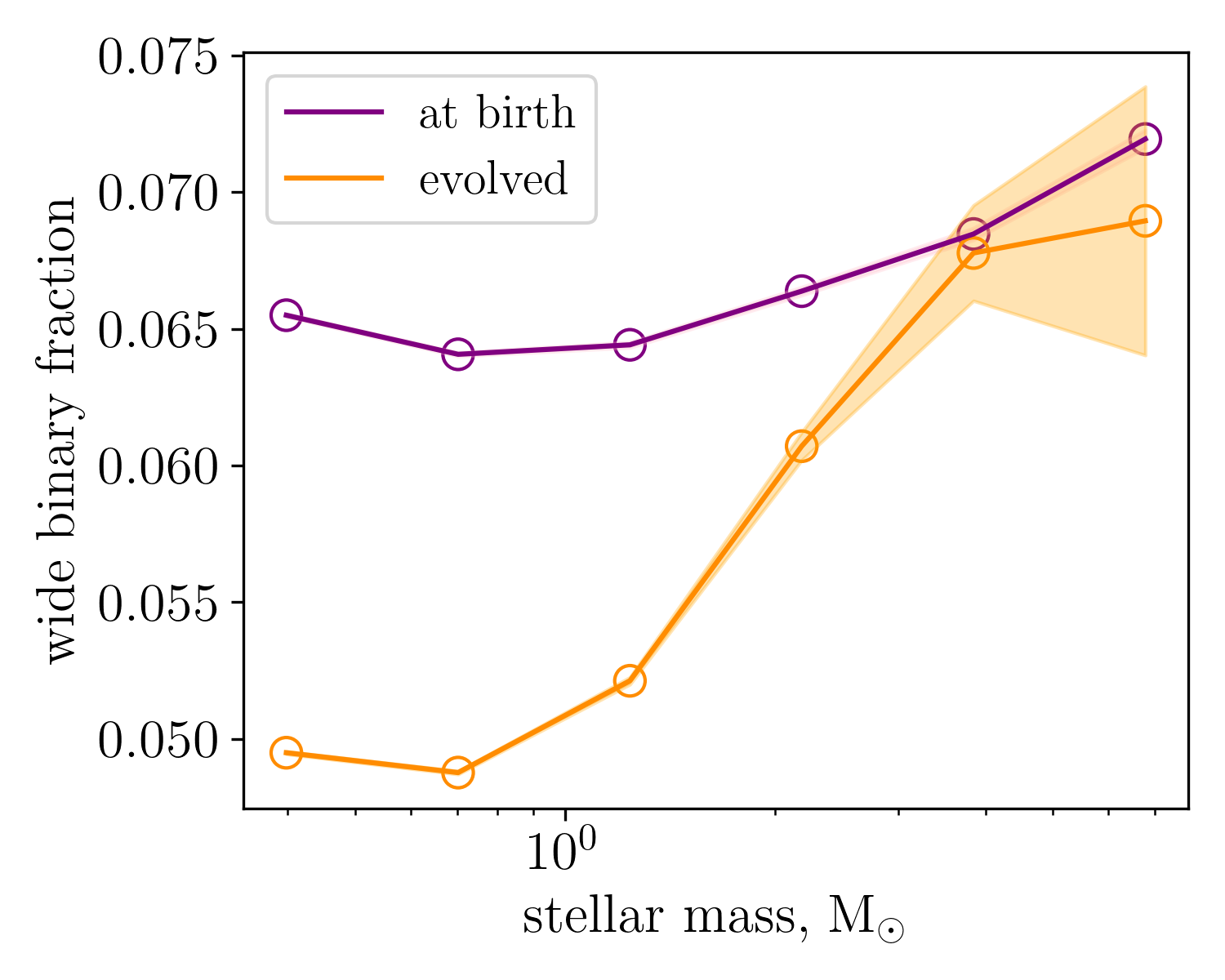}
    \caption{Mass dependence of the MS-MS binary fraction at birth (purple) and evolved to present day (orange). Left: the dependence of the binary fraction on mass assuming that the overall present-day binary fraction (including binaries of all separations) is 33\%. While at birth the binary fraction is independent of mass, some modest dependence is acquired as a result of evolution. The analytical model (black / grey) is described in the Appendix. Middle: only for wide binaries in our complete population synthesis model, assuming that the overall present-day wide-binary fraction (between 1000 and $10^4$ AU) is 2.4\%. Most of the mass dependence is acquired at birth due to the different period distributions of low-mass and high-mass binaries. Right: the present-day wide-binary fraction and the wide-binary fraction at birth in the model where the periods are drawn from the distribution of \citet{duqu91} regardless of mass. Shaded areas reflect Poisson errors in the finite stellar population synthesis sampling.}
    \label{fig:model_MS_MS}
\end{figure*}

Because we assign masses within the binary independently from each other (drawn from the \citealt{krou01} initial mass function), at birth the binary fraction is independent of mass. As the stellar population evolves, the binary fraction acquires some modest mass dependence, as illustrated in Fig.~\ref{fig:model_MS_MS} (left). Specifically, low-mass stars are on average significantly older than the high-mass stars, and therefore by the present epoch their companions are more likely to have evolved off of the main sequence than the companions of the high-mass stars, and therefore the binary fraction of low-mass stars is lower. For high-mass stars, the companions are drawn from the same \citet{krou01} mass distribution with very few massive companions, with few, if any, that have had time to evolve off of the main sequence, and therefore again because of their youth, the binary fraction is essentially the same as the binary fraction at birth. In the case of the birth binary fraction independent of mass, the present-day binary fraction can be calculated analytically as described in the Appendix, and the result of the numerical evaluation of the analytical integrals is shown in Fig.~\ref{fig:model_MS_MS} (left) in black / grey. 

The more observationally relevant regime for our investigation is that of wide binaries, with semi-major axes $10^3-10^4$ AU, and we ignore any projection effects. For the purpose of the wide binary fraction calculation, we imagine that any binary with a semi-major axis $<$100 AU is seen in the survey as a single star. The stars in the separation regime $100-1000$ AU contribute to the overall normalization of eq. (\ref{eq:bfraction}), but not to the numerator, which captures only the wide binaries. We show the resulting mass dependence at birth and at present in Fig.~\ref{fig:model_MS_MS}. We see that while there is some change of the mass dependence of the wide-binary fraction between birth and present day, qualitatively similar to the one described above (massive stars have their birth binary fraction, and the less massive stars evolve toward smaller binary fractions as they age), overall the mass dependence is much stronger than in the previous case, with the wide-binary fraction changing by a factor of 4 over a decade in mass, and its present-day shape largely reflects its shape at birth.

We therefore investigate what effects our population synthesis model assumptions have on the observed mass-dependent wide-binary fraction. In particular, in our original setup we assign two different period distributions to higher-mass binaries \citep{duqu91} and to the lower-mass binaries \citep{fisc92}. Specifically, as summarized by \citet{elba18}, period distribution is log-normal for both subsets of binaries, but for binaries where one of the stars has mass $>$0.75\Msun, the mean $\log (P/{\rm days})$ is 4.8 and the dispersion in this value is 2.3, and otherwise the mean and the dispersion are 4.1 and 1.3. In other words, binaries with at least one high-mass component have a period distribution which is shifted to longer periods and is significantly broader. This results in an enhancement of wide binary fraction at high masses that we see in Fig.~\ref{fig:model_MS_MS} (middle). 

To test this hypothesis, we also calculate the mass-dependent wide binary fraction drawing binary periods exclusively from the distribution of \citet{duqu91} regardless of the mass. We obtain a wide-binary fraction which varies much more weakly as a function of mass (Fig.~\ref{fig:model_MS_MS}, right). The mild mass-dependence seen in this figure is due to the mass-dependent conversion of the period distribution to the semi-major axis distribution necessary when we define our wide-binary fraction using $a<10$ AU cut for close binaries seen as singles and $a=10^3-10^4$ AU cut for wide binaries. 

We conclude that evolution effects alone result in a mild 20\% enhancement of the wide-binary fraction at high stellar masses, but these effects are not sufficient for explaining the steeply varying MS-MS wide binary fraction seen in the combination of our data and those of \citet{Moe2017} shown in Fig.~\ref{fig:WBF-MS1}. This increase in the wide-binary fraction as a function of mass largely reflects the binary fraction at birth. 

In these MS-MS binary fraction calculations, the overall fraction of stars in wide binaries with semi-major axes $10^3-10^4$ AU is set to be 2.4\%, an adjustable parameter directly related to the $N_{\rm singles}/N_{\rm binaries}$ in our stellar population synthesis model. This value sets the overall scale of the mass-dependent wide-binary fractions in Fig.~\ref{fig:model_MS_MS} in the middle and on the right. The resulting overall fraction of stars in binaries (i.e., counting binaries with any separations) is 33\%, which sets the scale for the left panel of Fig.~\ref{fig:model_MS_MS}. 

\subsection{White dwarf -- main sequence (WD-MS) binaries}

WD-MS binaries suffer one episode of major mass loss during their evolution. We consider a linear decline of mass as a function of time over a timescale $\tau_{\rm AGB}$ and the corresponding recoil given by eq. (\ref{eq:vkick}) parameterized by $v_{\rm asym}$ and explore variations in both of these parameters. The initial-to-final mass relationship, fixed to the one by \citet{Cummings2018}, is characterized by the mass-loss fraction that steeply rises with mass. The only other major adjustable parameter of the model is $N_{\rm singles}/N_{\rm binaries}$ which we calculate self-consistently so that the overall wide binary fraction of main-sequence stars is 2.4\%. 

\begin{figure*}
    \centering
    \includegraphics[height=7cm]{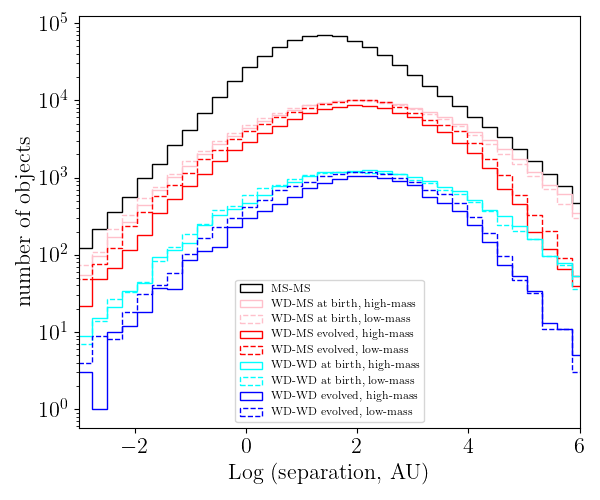}
    \includegraphics[height=7cm]{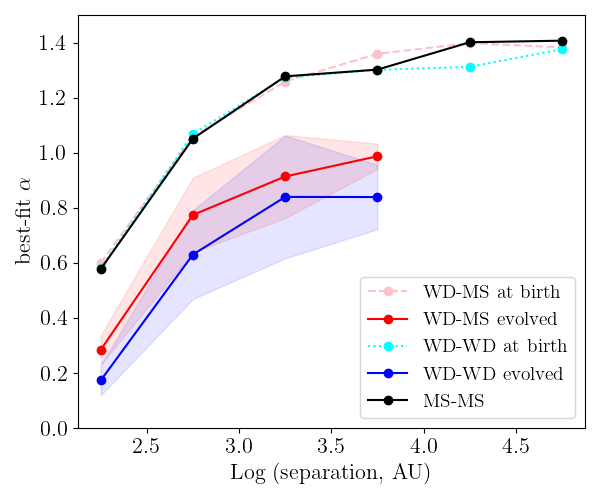}
    \caption{The results of our fiducial model with $v_{\rm asym}=0.25$ km s$^{-1}$ and $\tau_{\rm AGB}$ declining with mass from $2\times 10^6$ years at 1\Msun\ to $10^3$ years at 6\Msun. We show the distributions of separations (left) and eccentricities (right) of MS-MS binaries (black), WD-MS binaries (red), and WD-WD binaries (blue). For WD-MS and WD-WD binaries the values at birth (when they were still MS-MS binaries) are shown in pink and cyan, correspondingly. The high-mass vs low-mass boundary is at $m_{\rm WD}=0.64$\Msun, the median WD mass for evolved WD-MS pairs. Eccentricity indices are shown without any observational restrictions on the companion, but restricting the MS companions to our stated completeness range ($BP-RP>0.8$ mag and $M_G<10.5$ mag) produces no discernible change in the right panel.}
    \label{fig:model_SMA_ecc}
\end{figure*}

The distributions of semi-major axes for MS-MS, WD-MS and WD-WD binaries in our typical population synthesis model is shown in Fig.~\ref{fig:model_SMA_ecc}, left. On the short-separation end, the orbits of WD-MS binaries are larger than those of their MS-MS progenitors due to the adiabatic expansion of orbits, with $a(t)\propto 1/(m_1+m_2(t))$, as required by the conservation of Delaunay action $L$. Furthermore, since high-mass stars undergo higher fractional mass loss, the adiabatic expansion is slightly more pronounced for higher mass stars (solid lines) than for the lower mass stars (dashed lines). 

On the large-separation end, the decline of the separation distribution is significantly more steep for WD-MS and WD-WD binaries than for their progenitors and for the current MS-MS binaries, due to the preferential disruption of large-period orbits thanks to the recoil. Qualitatively, this steeper decline at high separations is consistent with the results of \citet{elba18}, although in their data this break in the semi-major axis distribution is observed at $10^{3.0-3.5}$ AU, whereas in all our models it occurs at $\le 10^3$ AU. For some models, there is a weak dependence of this break separation on the assumed value of $v_{\rm asym}$, with smaller kicks resulting in a break at slightly larger separations, but overall models with $v_{\rm asym}=0.25-1.5$ km s$^{-1}$ are qualitatively consistent with the data. We have also verified that setting $v_{\rm asym}=0$ results in a distribution of semi-major axes for WD-MS binaries that has the same shape as that of the MS-MS binaries on the long-period end, which is inconsistent with the data. Therefore, in agreement with the conclusions of \citet{elba18}, we find that models with recoils are strongly preferred to models with no recoil. 

The eccentricity distribution of WD-MS binaries predicted by our population synthesis model is shown in Fig.~\ref{fig:model_SMA_ecc}, right. The eccentricity distribution of WD-MS binaries is affected by three factors. First, due to the adiabatic expansion of the orbit during the mass loss, the binaries that end up at separations where we observe them must have started at smaller separations, where typical wide binary eccentricities are smaller \citep{Hwang2022ecc}. This lowers the eccentricities of the WD-MS binaries compared to the MS-MS binaries at the same separations. The median expansion of the orbit for final separations of the most observational interest to us ($10^3-10^4$ AU) is a factor of $\sim 2$, which is quite close to the median expected adiabatic expansion. The second effect is that the highest eccentricity orbits have the highest chance of getting disrupted during the mass loss and corresponding kick, removing these high-eccentricity binaries from the WD-MS population and further lowering the eccentricity of the remaining WD-MS binaries.

The third effect is that the orbital evolution of the surviving WD-MS binaries due to the recoil tends to increase the eccentricities. For the binaries at final separations $10^3-10^4$ AU, this is a small effect, which is insufficient to offset the decrease in eccentricities resulting from the former two effects. Thus the net eccentricities of the WD-MS binaries at $10^3-10^4$ AU in our population synthesis models are appreciably lower than those of the MS-MS binaries at the same separations. This is in good agreement with the observational result in Fig.~\ref{fig:alpha_sep}. 

\subsection{Mass dependence of the WD wide binary fraction}

Our observational results in Fig.~\ref{fig:wbf-mass} clearly indicate that the wide binary fraction for WD-MS and WD-WD binaries is a steeply declining function of mass, with the exception of low-mass ($\la 0.5$\Msun) WD-WD binary fraction. Low-mass white dwarfs are exclusively products of close stellar evolution \citep{Brown2022} because the main-sequence lifetime of their progenitors in the single stellar evolution production pathway would be longer than the age of the Universe. Therefore, all of the sources shown in Fig.~\ref{fig:wbf-mass} at low masses must be at least triple system -- a close binary with the low-mass white dwarf being the visible star, plus a distant companion which qualifies the object as a wide binary by our definition. Since our model does not include close binary evolution or hierarchical triples, we cannot comment on any of the results for these sources, so in the following discussion we focus exclusively on white dwarfs with $m_{\rm WD}>0.5$\Msun.  

Both the fractional mass loss and the coupled net velocity recoil are higher in the high-mass stars, thanks to the initial-to-final mass relation from \citet{Cummings2018}: at the initial mass 1\Msun, the star loses $\sim 40\%$ of its mass on the way to the white dwarf, whereas at the initial mass of 6\Msun\ it loses over 80\%. With $v_{\rm asym}=1$ km s$^{-1}$, the total physical recoil velocities -- when tied to the mass loss via momentum conservation -- range between $0.5$ and $1.8$ km s$^{-1}$. Therefore, at face value, it would seem that the declining mass dependence of the WD-MS and WD-WD wide binary fraction should be easy to explain since our model already includes these strong mass-dependent orbital disruption mechanisms. However, in practice reproducing the wide binary fraction of WD-MS binaries which declines as a function of mass proved difficult in most models that we have tried. While the mass-loss and the recoil are indeed more disruptive for more massive stars, the adiabatic expansion of the orbits is also stronger, and since the semi-major axis distribution is a declining function of separation, the high-mass progenitors originate at more populated separations, and this effect offsets their relatively greater chance of disruption. 

\begin{figure*}
    \centering
    \includegraphics[height=4.9cm]{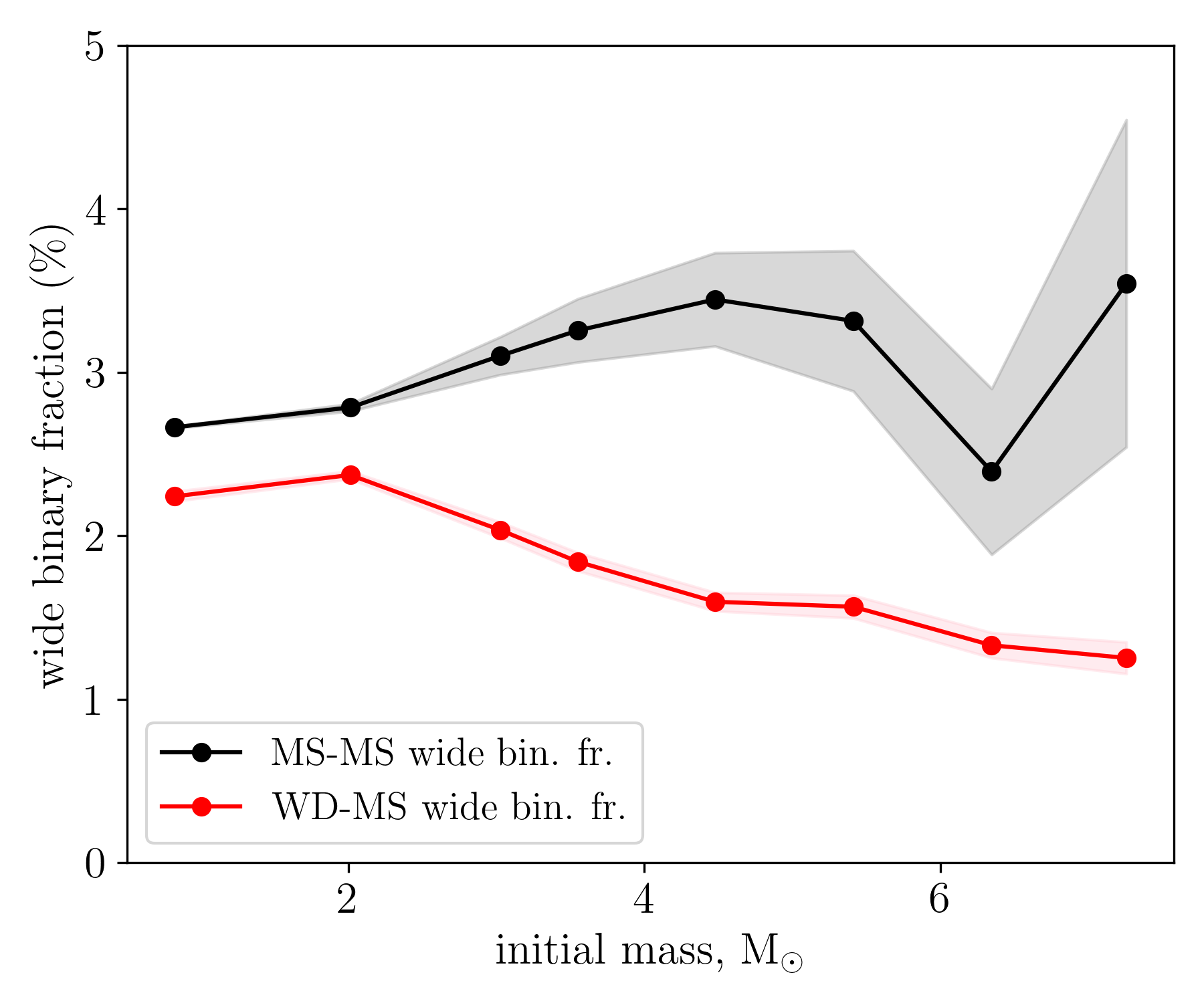}
    \includegraphics[height=4.9cm]{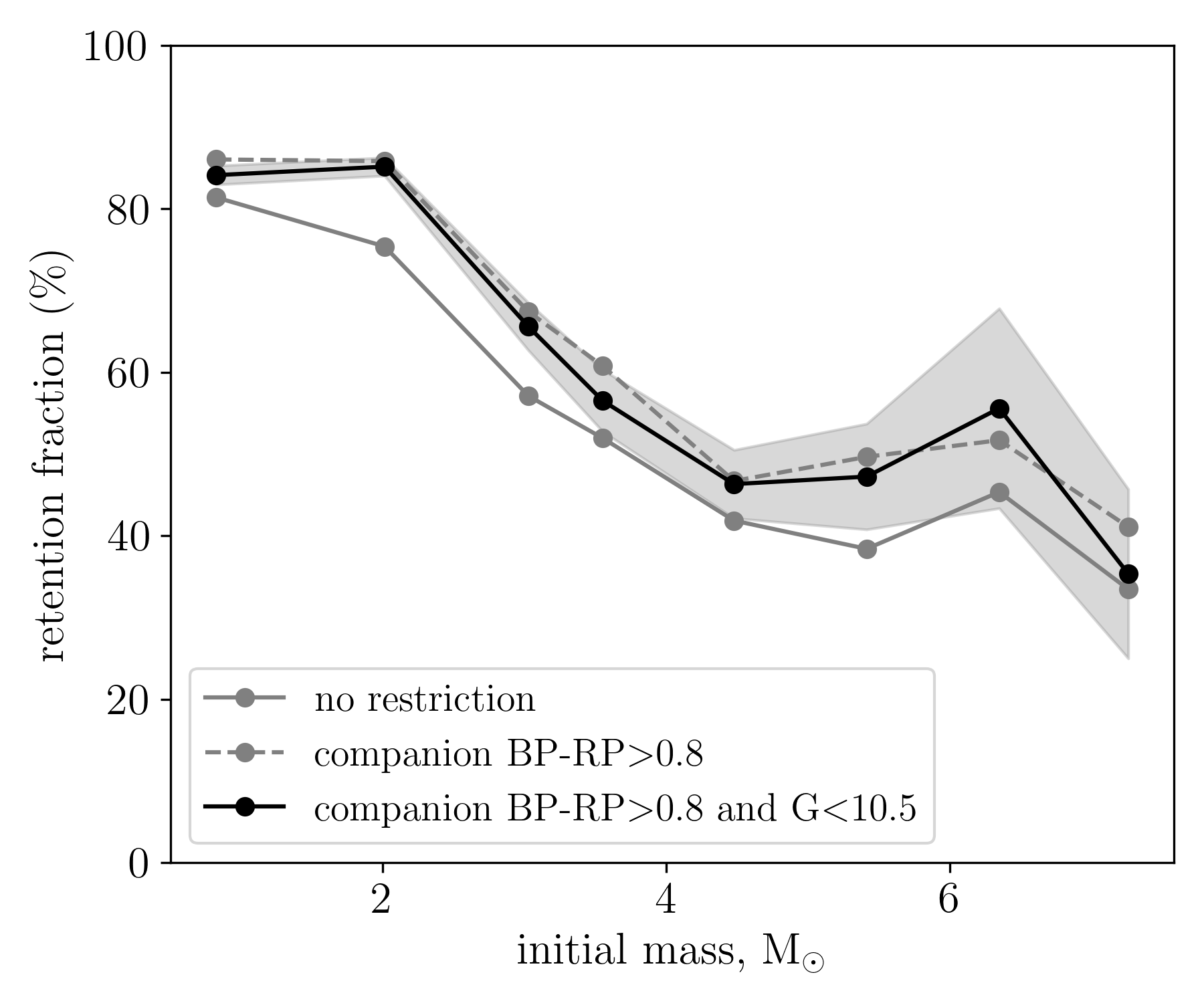}
    \includegraphics[height=4.9cm]{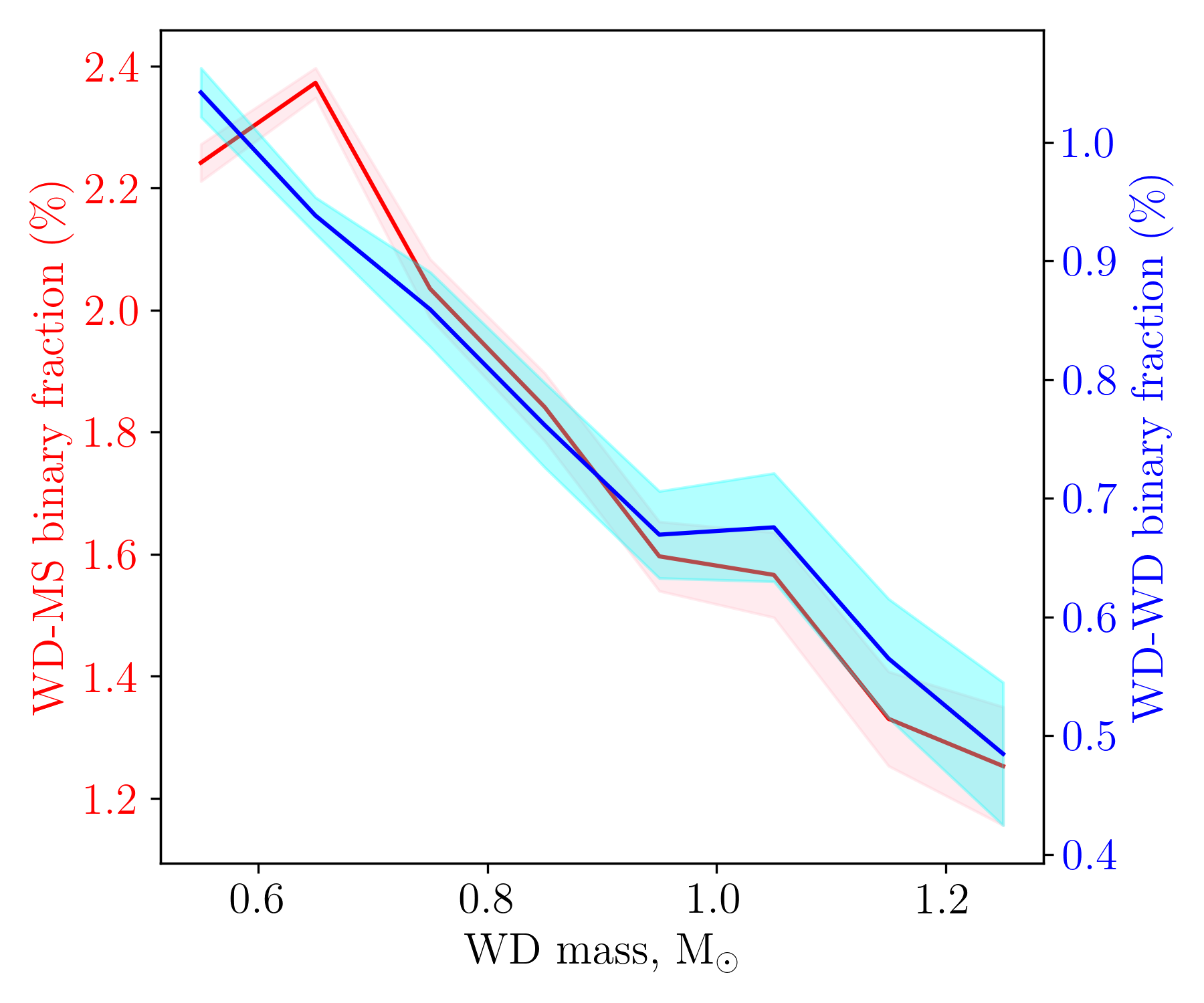}
    \caption{Mass-dependent MS-MS and WD-MS wide binary fractions (left), the retention rate (middle) and the WD-MS and WD-WD binary fractions (right) in the fiducial model with $v_{\rm asym}=0.25$ km s$^{-1}$ and a mass-dependent $\tau_{\rm AGB}$ which declines as a power-law function of mass from $2\times 10^6$ years at $m=1$\Msun\ to $10^3$ years at $m=6$\Msun. Observational restrictions ($BP-RP>0.8$ mag and $M_G<10.5$ mag) have been applied to the MS companions.}
    \label{fig:model_retention}
\end{figure*}

Thus two observational results are in tension with one another: the relatively lower eccentricities of the WD-MS binaries (compared to those of MS-MS binaries) are most naturally explained by the adiabatic expansion, but that same adiabatic expansion produces the wrong mass dependence of the wide binary fraction. The most successful model we have for simultaneously explaining these two observables involves a mass-dependent $\tau_{\rm AGB}$. Specifically, in Fig.~\ref{fig:model_retention} we show the mass-dependent wide binary fraction and the retention fraction for a model where $\tau_{\rm AGB}$, the characteristic timescale of the mass loss, declines as a power-law function of mass from $2\times 10^6$ years at $m=1$\Msun\ to $10^3$ years at $m=6$\Msun, and for masses outside of this range the timescale is fixed to the corresponding end values. This model still ensures that the bulk of the stars undergo relatively slow evolution which leads to adiabatic expansion of the orbits and with it, the relatively low eccentricities of the bulk of the WD-MS binaries are inherited from progenitors at somewhat lower separations. But for high mass stars, the evolution is quite fast and is not in the adiabatic regime for most of the wide binaries of interest, so the mass-dependent mass loss and recoil successfully lower the WD-MS wide binary fraction on the high-mass end. This effect can also be seen in Fig.~\ref{fig:model_SMA_ecc}, left, where at large separations the separation distribution of higher mass WD-MS and WD-WD binaries declines just a little more steeply than that of the lower mass WD-MS and WD-WD binaries. We have found that only with a mass-dependent $\tau_{\rm AGB}$ are we able to produce this effect. 

AGB and especially post-AGB evolution of stars remain among the more difficult stretches of stellar evolution to tackle theoretically \citep{vass94, bloe95a, bloe95b, weis09, mill16, dohe15}. The challenges include the extreme sensitivity of the relevant processes to metallicities and abundance patterns and resulting opacities, the necessity to include a wide range of relevant temporal scales, and numerical challenges associated with treating both radiative and convective transport \citep{weis09}. \citet{weis09} predict that low-mass stars ($<2$\Msun) end up at essentially their core masses by the end of the AGB phase, whereas massive stars ($M>2$\Msun) still retain much of their envelope at that point and have tens of per cent of their mass to lose in the post-AGB phase. Therefore, the relevant timescales are AGB timescale for low-mass stars and post-AGB timescales for high-mass ones. The timescale of post-AGB evolution varies by more than an order of magnitude among different calculations \citep{mill16}, but they all show post-AGB timescale declining steeply as a function of mass to $10^2-10^3$ years on the most massive end (4\Msun\ in his models). Therefore, at face value our scaling of the mass-loss timescale is qualitatively consistent with theoretical models by \citet{mill16}.

The existing observational constraints are in some tension with predictions from such models which predict that there should be stars losing mass at rates of $>10^{-3}$ \Msun\ year$^{-1}$. While initially the so-called OH/IR stars were considered to be candidates for such ``superwinds", the discovery of binary companions shaping their outflows have cast doubt on the high mass-loss estimates \citep{deci19}. Massive stars are rare and the evolutionary stage in question is short, and as a result probing the final significant mass loss episodes with direct observations is challenging. While in Fig.~\ref{fig:model_retention} we show our fiducial model with $\tau_{\rm AGB, massive}=10^3$ years on the most massive end, this is not the only acceptable model. The model with $\tau_{\rm AGB, massive}=10^4$ years on the massive end works well, but the one with $\tau_{\rm AGB, massive}=10^5$ years does not -- it does not produce a reasonable mass-dependent binary fraction because massive binaries expand adiabatically and yield overly abundant WD+MS binaries. Therefore, the observations and dynamical modeling presented here serve as independent useful constraints on the stellar evolutionary models and indicate that the characteristic mass loss timescale for massive stars should be $\la 10^4$ years.

\subsection{Eccentricities of WD+MS binaries}

The model with $\tau_{\rm AGB}$ being a steep function of mass makes another testable prediction, that the eccentricities of massive WD binaries with an MS star (massive WD-MS) and those of low-mass WD binaries with an MS star (low-mass WD-MS) should behave differently. We fix $\tau_{\rm AGB, low-mass}=2\times 10^6$ years on the low-mass end and vary $\tau_{\rm AGB, massive}$ on the high-mass end. For $\tau_{\rm AGB, massive}=10^5$ years, all binaries at $a<10^4$ AU are in the adiabatic regime, and the massive WD-MS binaries have lower eccentricities than low-mass WD-MS binaries because for the former, the fractional mass loss is greater and they adiabatically expand over a larger range of $a$, putting initially lower eccentricity orbits into the observed range of separations. In contrast, for $\tau_{\rm AGB, massive}\la 10^4$ years massive WD-MS binaries at $a=10^3-10^4$ AU should have higher eccentricities because none of them got there adiabatically. The models predict a big difference in $\alpha$ at $a=10^3-10^4$ AU: $\alpha=1.2-1.4$ vs $0.6-0.8$ for massive WD-MS binaries when $\tau_{\rm AGB, massive}=10^3$ years and $10^5$ years, correspondingly, with massive WDs defined as those with $m_{\rm WD}>0.7$\Msun\ (birth mass $>3$\Msun). In the models the effect on the eccentricity is noticeable when massive WD-MS binaries are defined as those with $m_{\rm WD}>0.65$\Msun\ (birth mass $>2$\Msun) and increases with mass.

\begin{figure}
	\centering
	\includegraphics[width=1\linewidth]{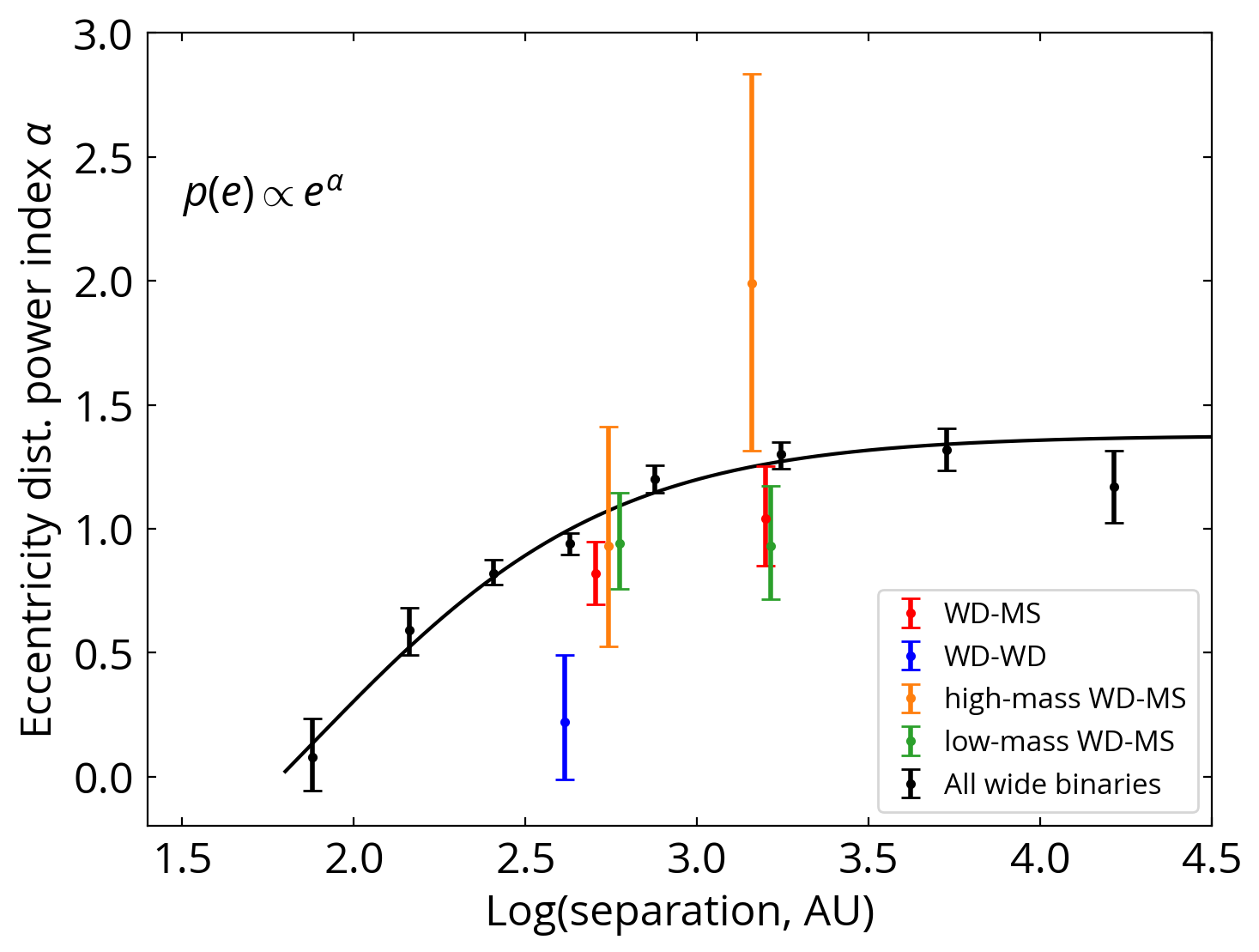}
	\caption{Same as Fig.~\ref{fig:alpha_sep}, right, but now with the WD-MS sample split into high-mass and low-mass WD-MS binaries based on the mass of the WD ($0.8-1.4$\Msun\ and $0.5-0.8$\Msun, correspondingly). At relatively low separations $a=10^{2.75}$ AU, massive WD-MS binaries have the same or (depending on the exact mass cuts) somewhat lower eccentricities than the low-mass ones. At higher separations, massive WD-MS binaries are noticeably more eccentric. Our models suggest that this is a result of a significantly shorter evolutionary scale $\tau_{\rm AGB}$ for massive stars than for low-mass stars, and that $\tau_{\rm AGB, massive}$ must be comparable to the orbital timescale at $a=10^{2.75}$ AU, a few thousand years.}
	\label{fig:highmass_ecc}
\end{figure}

In the observed sample, we split the WD-MS sample of binaries into those with $m_{\rm WD}=0.5-0.8$\Msun\ and those with $m_{\rm WD}=0.8-1.4$\Msun\ and show the results of the eccentricity measurements in Fig.~\ref{fig:highmass_ecc}. Although the statistics become poor for the high-mass bin, we see a trend that qualitatively agrees with the theoretical expectations: at lower separations, high-mass WD-MS binaries have same eccentricities as the low-mass ones, and at higher separations, significantly higher. The exact values of $\alpha$ are sensitive to the WD mass cutoff, but for all options we have tried, at $a=10^{2.75}$ AU the eccentricities of high-mass WD+MS are similar to or lower than those of low-mass WD+MS binaries, but are always significantly higher at $a=10^{3.25}$ AU. The low number of sources is not the only challenge in interpreting these results -- the initial eccentricity distribution of MS-MS binaries from \citet{Hwang2022ecc} which initiates our theoretical calculations is not well known for high-mass MS stars. Thus it is not known whether the observed differences in the WD-MS eccentricities as a function of WD mass may be due to their unknown birth eccentricities or due to the evolutionary effects modeled here.

If we assume that the evolutionary effects dominate, then we can draw one further conclusion from the observed results, especially at $a=10^{2.75}$ AU where the massive WD-MS binaries are observed to have similar or slightly lower eccentricities than the low-mass ones. The relative eccentricities of the two samples are particularly sensitive to $\tau_{\rm AGB, massive}$ at these separations, because the orbital period -- initially 1400 years and eventually 9700 years (for a $4+1$\Msun\ initial MS-MS binary that expands and loses mass) straddles the transition from the adiabatic to impulsive regime for $\tau_{\rm AGB, massive}=10^{3-4}$ years. Models predict that eccentricities should be the same for the two WD mass bins at $\tau_{\rm AGB, massive}=10^3$ years, but at $\tau_{\rm AGB, massive}=10^4$ years massive WD-MS are expected to have significantly lower eccentricities. The observed result is, within the error bars, consistent with both and is in between these two scenarios, and the observed mass dependence of $\alpha$ is the opposite of the expectation for $\tau_{\rm AGB}=10^5$ years. 

There is another interesting effect potentially seen both in observations and simulations for high-mass WD+MS binaries. As mentioned in Sec. \ref{sec:pop}, high-eccentricity orbits are preferentially disrupted, so the final eccentricity distribution in the model only resembles a power-law at $e\la 0.8$, with a paucity of higher-eccentricity orbits. Interestingly, the distribution of the $v-r$ angles for the massive WD+MS binaries at $a=1000-3000$ AU as a peak at $20^{\rm o}$ and a relative dearth at $0^{\rm o}-10^{\rm o}$. This may indicate that orbits with eccentricities $>0.94$ \citep{Hwang2022ecc} are lacking. This may be in qualitative agreement with simulations -- quantitative agreement is not necessarily expected because many aspects of the simulations, such as the exact $m(t)$ during mass loss, are not very faithfully reproduced.

\subsection{Effects of hierarchical triples}

Many stars are in hierarchical triple systems \citep{Law2010, Tokovinin2017}, and the incidence of triples may be a strongly increasing function of stellar mass \citep{Duchene2013}. Stars in multiple systems are most often arranged hierarchically, so that our wide binary sample could be contaminated by hierarchical triples and quadruples with tight binaries in one or both visible components of the apparent binary. Our selection of stars purposefully removes many such triples, thanks to the \texttt{ruwe}$<1.4$ cut on the MS stars and to the astrometric quality selection cuts imposed on WDs in \citet{gent21}, specifically \texttt{astrometric\_sigma5d\_max} $< 1.5$ \texttt{or} (\texttt{ruwe}$\le 1.1$ \texttt{and} \texttt{ipd\_gof\_harmonic\_amplitude} $<1$). In this section we qualitatively discuss how triples and the potential mass dependence of their incidence might impact our results.

The observed mass-dependent wide binary fraction for WD-MS and WD-WD binaries (Fig.~\ref{fig:wbf-mass}) shows a break at 0.6 \Msun, which is likely due to the low-mass white dwarfs being products of close stellar evolution. These wide binaries are therefore hierarchical triple systems with a very close binary whose luminosity is dominated by the low-mass white dwarf and a distant companion which puts it in our wide binary sample. Since our model does not include triple systems or close stellar binaries, we cannot model these sources. First we ensure that the mass dependence of the effects shown in Fig.~\ref{fig:wbf-mass} is robust to any changes to target selection. \citet{gent21} parent sample of WDs may be incomplete at the low-mass end as they sought to avoid WD+MS close binaries, but this mass incompleteness would affect the selection of singles and the selection of WDs in wide binaries in the same way, so it does not affect the binary fraction. We then experiment with a different range of binary separations included in the measurement, with a different parallax cut, and with setting a different \texttt{ruwe} cut on the MS companions, including increasing it all the way to \texttt{ruwe}$<40$. None of these modifications result in a qualitative change in the mass dependence of the binary fractions. For example, relaxing the \texttt{ruwe} cut to $<40$ leads to an increase of the WD-MS binary fraction by 10\% (because this cut lets in some hierarchical triples) with no changes in the functional shape.

Qualitatively, the binary fraction of the extremely low mass WDs should be lower (as observed) because triples are rarer than binaries and because such objects have likely undergone common envelope evolution, with potentially high fractional mass loss and high kicks that are more likely to have disrupted them. Some triples with the low-mass WD+WD close binaries accompanied by the distant MS companion may be filtered by the \citet{gent21} astrometric quality cuts. Those that do remain in our sample likely have orbital periods that are significantly shorter than those that can be probed by {\it Gaia} astrometric noise cuts. The majority of extremely low mass white dwarfs in the survey by \citet{brow20} have periods $<1$ day, which, for a binary at 200 pc, translates to orbital separations of $<0.1$ mas, out of reach for detection via a \texttt{ruwe} flag for faint sources by about 1 dex \citep{Belokurov2020}.

Another effect we do not take into account in our model is the mergers of close low-mass WD-WD binaries which form massive white dwarfs, with some authors suggesting that a large fraction of all white dwarfs with $m_{\rm WD}\ga 0.8$ \Msun\ may be due to such mergers (e.g., \citealt{kili18}). If such mergers are indeed a significant source of massive white dwarfs, then this population would further lower our model wide binary fraction at high mass and improve the agreement between the model and observations. 

We now examine the hierarchical triples where the close binary does not undergo mass exchanges and mergers, i.e., remains detached throughout its evolution. To elucidate the effects of the triple fraction on the observed mass dependence, we consider an extreme toy model in which wide binaries that contain a massive star at birth are all turned into hierarchical triples. We consider two scenarios. In the first, the massive star ($>3$\Msun) is by itself, but the wide binary companion is a close MS+MS binary, where the third companion is drawn from the \citet{krou01} mass function but cut off at 3\Msun. At face value, one might expect that all the effects of stellar mass loss that we have discussed would be suppressed because the fractional mass loss becomes smaller with an addition of another companion. In the simulation we find instead that this effect is offset and superseded by the fact that some of the closer triples might appear as relatively massive singles after the most massive star evolves into a WD, and the retention fraction as a function of mass is somewhat steeper than that shown in Fig.~\ref{fig:model_retention}, middle, i.e., it agrees better with observations. There is no discernible effect on any other observables such as eccentricities.

Another scenario is that of the massive star in a close binary with an MS star with a distant MS companion. In this case after the massive star evolves to a WD, the system appears as a wide MS+MS binary, since a WD quickly becomes invisible next to its close MS companion at our $M_G<10.5$ mag brightness cut. Therefore the apparent retention fraction is significantly reduced at high masses. Overall, although we do not fully and self-consistently treat triple systems in our simulations, it seems that their contribution is essential to explain the mass dependence of the retention fraction at low masses and they may aid in improving the agreement between the retention fraction seen in the theoretical models (Fig.~\ref{fig:model_retention}) and observations (Fig.~\ref{fig:retention}) on the high-mass end.

\section{Discussion and conclusions}
\label{sec:conclusions}

In this paper we present novel measurements of the orbital statistics of wide binaries containing white dwarfs and use these measurements to connect stellar evolutionary processes (specifically, the mass loss that the stars undergo on their way from the main sequence to the white dwarf phase) and their orbital dynamics. This work builds on that of \citet{elba18} who measured the distributions of semi-major axes of MS-MS, WD-MS and WD-WD wide binaries, found them to be significantly discrepant, and developed stellar population models that strongly suggest that white dwarfs experience a typical kick of $\sim 0.75$ km s$^{-1}$ at birth, perhaps due to the anisotropic mass loss. This kick results in a preferential disruption of the widest binaries whose orbital velocities are similar to the kick and steepens the semi-major axis distribution, as observed. 

We present two new observational results. First, we find that the wide binary fraction for binaries containing a white dwarf (WD-MS and WD-WD binaries) is a steeply declining function of white dwarf mass, whereas in the same range of separations ($10^3-10^4$ AU) the progenitors of these binaries experience a strong opposite trend, with the binary fraction increasing as a function of mass roughly two-fold (Fig.~\ref{fig:WBF-MS1}). Second, we find (Fig.~\ref{fig:alpha_sep}) that the eccentricities of WD-MS binaries and especially those of WD-WD binaries are significantly lower than those of the MS-MS binaries at the same separation ($10^{2.5}-10^3$ AU).

We expand the theoretical model of \citet{elba18} in order to explain all these results simultaneously. The most significant alteration is that instead of an instantaneous mass loss episode and a velocity kick at the end of the main sequence evolution we consider a range of timescales $\tau_{\rm AGB}$ which may be shorter than, longer than, or comparable to the orbital periods of interest. Another alteration is to tie the amount of velocity recoil to the fractional mass loss. Finally, the eccentricities of binaries at birth are drawn from different distributions depending on their birth separation \citep{Hwang2022ecc}.

We find that the decreased eccentricities of WD-MS and WD-WD binaries in comparison to those of the MS-MS binaries are most naturally explained by the adiabatic expansion of the orbits during mass loss, when the semi-major axis is inversely proportional to the (decreasing) total mass of the binary, as required by the conversation of an adiabatic invariant of the problem. Such binaries started their MS-MS lives at lower separations where typical eccentricities are lower, and even though in the presence of the recoil the eccentricities of individual systems evolve, the adiabatic expansion effect appears to dominate. In order for this mechanism to succeed, the timescale for mass loss for the majority of white dwarfs should be significantly longer than the orbital timescale at $10^{2.5}$ AU, that is $\tau_{\rm AGB}\gg 3,000$ years. 

In contrast, the steeply declining WD-MS and WD-WD binary fractions as a function of mass require that the high-mass binaries cannot be in the adiabatically expanding regime: if they were, then the more massive binaries would have started their lives at lower separations than the less massive binaries, and thanks to the steeply declining period distribution of binaries even their larger fractional mass loss would be insufficient to explain the observed binary fraction mass dependence. Therefore, for the massive stars the evolutionary timescale must be comparable to or smaller than the orbital timescale at $10^3$ AU, that is $\tau_{\rm AGB}\la 10,000$ years. Therefore, explaining both observations requires a strongly mass-dependent $\tau_{\rm AGB}$ timescale for mass loss; not one model with a constant $\tau_{\rm AGB}$ was successful. Our fiducial model has $\tau_{\rm AGB}$ declining as a power-law from $2\times 10^6$ years at 1\Msun\ to $10^3$ years at 6\Msun. While this model does result in a decline of the binary fraction with mass, it is not quite as dramatic as observed: the WD-MS wide binary fraction, the retention fraction and the WD-WD wide binary fraction all decline by a modest factor of 2 at $m_{\rm WD}>0.6$\Msun, not by a factor of $>$6 (cf. Fig.~\ref{fig:WBF-MS1} and Fig.~\ref{fig:model_retention}, left; Fig.~\ref{fig:retention} and Fig.~\ref{fig:model_retention}, middle; and Fig.~\ref{fig:wbf-mass} and Fig.~\ref{fig:model_retention}, right). 

The model with $\tau_{\rm AGB, massive}=10^4$ years on the massive end performs similarly to the one with $10^3$ years. Its binary fraction mass dependence is less steep and is therefore in greater tension with observations than the fiducial model with $10^3$ year mass loss timescale for massive stars, but may reproduce the eccentricities of massive WD-MS binaries slightly better, although the observational statistics are too poor to tell unambiguously. In contrast, the model with $\tau_{\rm AGB, massive}=10^5$ year timescale fails to reproduce the declining binary mass dependence and the eccentricities as a function of WD mass altogether. The results are significantly less sensitive to the specific value of the velocity recoil than to the choice of $\tau_{\rm AGB}$. Our fiducial model has recoil in the range $0.25-1$ km s$^{-1}$, depending on the mass, but increasing these values by a factor of a few still results in qualitatively reasonable results. Finally, we confirm the results of \citet{elba18} -- that models with no recoil produce semi-major axis distributions that are significantly flatter than observed at large separations, essentially consistent with the shape of the separation distribution of these binaries at birth (shown in pink and cyan in Fig. \ref{fig:model_SMA_ecc}, left). 

\citet{Pham2024} develop a model of white dwarf accretion from the Oort-like cloud of comets to explain the observed widespread metal pollution on the surfaces of old white dwarfs. Their calculations are built on the results of \citet{OConnor2023} who demonstrate that a significant fraction of such exo-Oort cloud would survive the mass-loss stage of stellar evolution if the kick is $\sim 1$ km/sec and if the timescale for the mass-loss phase is $10^4-10^5$ years. The wide binary statistics presented here demonstrate that potentially both of these values -- the kick and the timescale -- are strongly dependent on mass, and if this is indeed the case, then for the most massive stars it may be hard to retain the exo-Oort cloud. Therefore, in the scenario proposed here one might expect that high-mass white dwarfs are significantly less likely to be polluted by their exo-Oort clouds than low-mass white dwarfs.

\section*{Availability of data and software}

All data used in this paper are from the public {\it Gaia} archive. The complete code used in the theoretical model is available at \url{https://github.com/zakamska/binaries} with a frozen repository in \citet{zenodo2025}.

\begin{acknowledgments}

The authors are grateful to Nicole Crumpler, Kareem El-Badry, Chris Hamilton, Dang Pham, Scott Tremaine and the anonymous referee for useful discussions. N.L.Z. is supported at the IAS by the J. Robert Oppenheimer Visiting Professorship and the Bershadsky Fund.

This work has made use of data from the European Space Agency (ESA) mission Gaia (\url{https://www.cosmos.esa.int/gaia}), processed by the Gaia Data Processing and Analysis Consortium (DPAC, \url{https://www.cosmos.esa.int/web/gaia/dpac/consortium}). Funding for the DPAC has been provided by national institutions, in particular the institutions participating in the Gaia Multilateral Agreement.

\end{acknowledgments}

\facilities{Gaia}

\software{\texttt{Astropy} \citep{Astropy2013, Astropy2018, Astropy2022}}

\appendix
\restartappendixnumbering
\setcounter{figure}{0}

\section{Mass dependence of the wide binary fraction as a function of time}

\citet{Moe2017} demonstrate that the fraction of stars in wide MS-MS binaries increases with mass. This is also seen in our data (Fig.~\ref{fig:WBF-MS1}). In this Appendix, we explore quantitatively how much of this increase can be produced due to the aging of the binary population alone as opposed to being required to be set at birth. 

Indeed, let us assume that the masses of the stars in the binary are independently drawn from the same mass distribution \citep{Moe2017}. A typical low-mass star is significantly older than a typical high-mass star. Therefore, a companion to a low-mass star is significantly more likely than a companion to a high-mass star to have evolved off of the main sequence during the life time of the binary, eliminating it from the MS-MS binary budget. We can make a further prediction that for stellar masses that have MS ages longer than the Hubble time -- i.e., $m\la 0.9$\Msun\ -- the binary mass fraction should be independent of mass, since the age distribution of the companions to these stars does not depend on their mass. 

We calculate the magnitude of this effect both analytically and using a binary population synthesis simulation. In the analytical calculation, $p(m){\rm d}m$ is the normalized mass function of single stars ($\int p(m){\rm d}m=1$) and $\tau(m)$ is the MS lifetime at mass $m$ \citep{lame17}. The star formation rate (SFR: total number of single stars born per unit time) is constant between $0$ and $t_0=12$ Gyr, and the binary fraction at birth $f_{\rm in}$ is independent of mass, but is unknown {\it a priori}. If a star is born at $t'\in (0, t_0)$, then it is on the MS at the present time if its MS lifetime is greater than the elapsed time, $\tau(m)>t_0-t'$, otherwise it has evolved away. This condition can be mathematically formalized as 
\begin{equation}
    h(m,t',t_0)=H(\tau(m)-(t_0-t')),
\end{equation}
where $H(x)$ is the Heaviside function, so that $h(m,t',t_0)$ is $=1$ if the star is on the MS and $=0$ otherwise. 

This allows us to formally write down the number of singles between $m$ and $m+{\rm d}m$ that are still on the main sequence at the present time:
\begin{equation}
    N_{\rm s}(m){\rm d}m={\rm SFR} p(m){\rm d}m\int_{t'=0}^{t_0}h(m,t',t_0){\rm d}t'.
\end{equation}
Similarly, the number of stars in the same mass range which are still on the main sequence and are in binaries with another MS star is
\begin{equation}
    N_{\rm bb}(m){\rm d}m=2 f_{\rm in}{\rm SFR} p(m){\rm d}m\int_{t'=0}^{t_0}\int_{m_2}p(m_2)h(m,t',t_0)h_2(m_2,t',t_0){\rm d}m_2{\rm d}t'.
\end{equation}
Some stars start off as MS-MS binaries, but at the present day one star has evolved off the main sequence, so these objects would be counted by an observer as singles. The number of these objects is
\begin{equation}
    N_{\rm bs}(m){\rm d}m=2 f_{\rm in}{\rm SFR} p(m){\rm d}m\int_{t'=0}^{t_0}\int_{m_2}p(m_2)h(m,t',t_0)\left(1-h_2(m_2,t',t_0)\right){\rm d}m_2{\rm d}t'.
\end{equation}
The overall currently observed MS-MS binary fraction $f_0$ can then be used to calculate the unknown binary fraction at birth $f_{\rm in}$: 
\begin{equation}
    f_0=\frac{\int_m N_{\rm bb}(m){\rm d}m}{\int_m N_{\rm s}(m){\rm d}m + \int_m N_{\rm bs}(m){\rm d}m + \int_m N_{\rm bb}(m){\rm d}m}.
\end{equation}
Finally, the mass-dependent binary fraction is then
\begin{equation}
    f(m)=\frac{N_{\rm bb}(m)}{N_{\rm s}(m)+N_{\rm bs}(m)+N_{\rm bb}(m)}.
    \label{eq:ap_fm}
\end{equation}
For some simple distributions these integrals can be evaluated analytically. For example we have checked that for a Bernoulli mass distribution (low-mass stars with mass $M_1$ are born with probability $p$ and high-mass stars with mass $M_2$ are born with probability $1-p$) the results agree with the expectation. For other distributions, the integrals in eq. (\ref{eq:ap_fm}) need to be evaluated numerically. Standard domain-binning schemes prove to be too computationally demanding, so we use the Monte Carlo integrator \texttt{MCINT}\footnote{\url{https://pypi.org/project/mcint/}}. An example resulting binary fraction as a function of mass is shown in Fig.~\ref{fig:model_MS_MS}, left. 

\bibliography{stars}{}

\begin{thebibliography}{}
\expandafter\ifx\csname natexlab\endcsname\relax\def\natexlab#1{#1}\fi

\bibitem[{{Abt} \& {Levy}(1976)}]{abt76}
{Abt}, H.~A., \& {Levy}, S.~G. 1976, \apjs, 30, 273

\bibitem[{{Astropy Collaboration} {et~al.}(2013){Astropy Collaboration},
  Robitaille, Tollerud, Greenfield, Droettboom, Bray, Aldcroft, Davis,
  Ginsburg, Price-Whelan, Kerzendorf, Conley, Crighton, Barbary, Muna,
  Ferguson, Grollier, Parikh, Nair, G{\"{u}}nther, Deil, Woillez, Conseil,
  Kramer, Turner, Singer, Fox, Weaver, Zabalza, Edwards, {Azalee Bostroem},
  Burke, Casey, Crawford, Dencheva, Ely, Jenness, Labrie, Lim, Pierfederici,
  Pontzen, Ptak, Refsdal, Servillat, \& Streicher}]{Astropy2013}
{Astropy Collaboration}, Robitaille, T.~P., Tollerud, E.~J., {et~al.} 2013,
  \aap, 558, A33

\bibitem[{{Astropy Collaboration} {et~al.}(2018){Astropy Collaboration},
  Price-Whelan, Sipocz, Gunther, Lim, Crawford, Conseil, Shupe, Craig,
  Dencheva, Ginsburg, VanderPlas, Bradley, P{\'{e}}rez-Su{\'{a}}rez,
  de~Val-Borro, Aldcroft, Cruz, Robitaille, Tollerud, Ardelean, Babej,
  Bachetti, Bakanov, Bamford, Barentsen, Barmby, Baumbach, Berry, Biscani,
  Boquien, Bostroem, Bouma, Brammer, Bray, Breytenbach, Buddelmeijer, Burke,
  Calderone, Rodr{\'{i}}guez, Cara, Cardoso, Cheedella, Copin, Crichton,
  D{\'{A}}vella, Deil, Depagne, Dietrich, Donath, Droettboom, Earl, Erben,
  Fabbro, Ferreira, Finethy, Fox, Garrison, Gibbons, Goldstein, Gommers, Greco,
  Greenfield, Groener, Grollier, Hagen, Hirst, Homeier, Horton, Hosseinzadeh,
  Hu, Hunkeler, Ivezi{\'{c}}, Jain, Jenness, Kanarek, Kendrew, Kern,
  Kerzendorf, Khvalko, King, Kirkby, Kulkarni, Kumar, Lee, Lenz, Littlefair,
  Ma, Macleod, Mastropietro, McCully, Montagnac, Morris, Mueller, Mumford,
  Muna, Murphy, Nelson, Nguyen, Ninan, N{\"{o}}the, Ogaz, Oh, Parejko, Parley,
  Pascual, Patil, Patil, Plunkett, Prochaska, Rastogi, Janga, Sabater,
  Sakurikar, Seifert, Sherbert, Sherwood-Taylor, Shih, Sick, Silbiger,
  Singanamalla, Singer, Sladen, Sooley, Sornarajah, Streicher, Teuben, Thomas,
  Tremblay, Turner, Terr{\'{o}}n, van Kerkwijk, de~la Vega, Watkins, Weaver,
  Whitmore, Woillez, \& Zabalza}]{Astropy2018}
{Astropy Collaboration}, Price-Whelan, A.~M., Sipocz, B.~M., {et~al.} 2018,
  \aj, 156, 123

\bibitem[{{Astropy Collaboration} {et~al.}(2022){Astropy Collaboration},
  {Price-Whelan}, {Lim}, {Earl}, {Starkman}, {Bradley}, {Shupe}, {Patil},
  {Corrales}, {Brasseur}, {N{\"o}the}, {Donath}, {Tollerud}, {Morris},
  {Ginsburg}, {Vaher}, {Weaver}, {Tocknell}, {Jamieson}, {van Kerkwijk},
  {Robitaille}, {Merry}, {Bachetti}, {G{\"u}nther}, {Aldcroft},
  {Alvarado-Montes}, {Archibald}, {B{\'o}di}, {Bapat}, {Barentsen},
  {Baz{\'a}n}, {Biswas}, {Boquien}, {Burke}, {Cara}, {Cara}, {Conroy},
  {Conseil}, {Craig}, {Cross}, {Cruz}, {D'Eugenio}, {Dencheva}, {Devillepoix},
  {Dietrich}, {Eigenbrot}, {Erben}, {Ferreira}, {Foreman-Mackey}, {Fox},
  {Freij}, {Garg}, {Geda}, {Glattly}, {Gondhalekar}, {Gordon}, {Grant},
  {Greenfield}, {Groener}, {Guest}, {Gurovich}, {Handberg}, {Hart},
  {Hatfield-Dodds}, {Homeier}, {Hosseinzadeh}, {Jenness}, {Jones}, {Joseph},
  {Kalmbach}, {Karamehmetoglu}, {Ka{\l}uszy{\'n}ski}, {Kelley}, {Kern},
  {Kerzendorf}, {Koch}, {Kulumani}, {Lee}, {Ly}, {Ma}, {MacBride}, {Maljaars},
  {Muna}, {Murphy}, {Norman}, {O'Steen}, {Oman}, {Pacifici}, {Pascual},
  {Pascual-Granado}, {Patil}, {Perren}, {Pickering}, {Rastogi}, {Roulston},
  {Ryan}, {Rykoff}, {Sabater}, {Sakurikar}, {Salgado}, {Sanghi}, {Saunders},
  {Savchenko}, {Schwardt}, {Seifert-Eckert}, {Shih}, {Jain}, {Shukla}, {Sick},
  {Simpson}, {Singanamalla}, {Singer}, {Singhal}, {Sinha}, {Sip{\H{o}}cz},
  {Spitler}, {Stansby}, {Streicher}, {{\v{S}}umak}, {Swinbank}, {Taranu},
  {Tewary}, {Tremblay}, {de Val-Borro}, {Van Kooten}, {Vasovi{\'c}}, {Verma},
  {de Miranda Cardoso}, {Williams}, {Wilson}, {Winkel}, {Wood-Vasey}, {Xue},
  {Yoachim}, {Zhang}, {Zonca}, \& {Astropy Project Contributors}}]{Astropy2022}
{Astropy Collaboration}, {Price-Whelan}, A.~M., {Lim}, P.~L., {et~al.} 2022,
  \apj, 935, 167

\bibitem[{Belokurov {et~al.}(2020)Belokurov, Penoyre, Oh, Iorio, Hodgkin,
  Evans, Everall, Koposov, Tout, Izzard, Clarke, \& Brown}]{Belokurov2020}
Belokurov, V., Penoyre, Z., Oh, S., {et~al.} 2020, \mnras, 496, 1922

\bibitem[{{Belyaev} \& {Rafikov}(2010)}]{bely10}
{Belyaev}, M.~A., \& {Rafikov}, R.~R. 2010, \apj, 723, 1718

\bibitem[{{Binney} \& {Tremaine}(2008)}]{binn08}
{Binney}, J., \& {Tremaine}, S. 2008, {Galactic Dynamics: Second Edition}
  (Princeton University Press)

\bibitem[{{Bloecker}(1995{\natexlab{a}})}]{bloe95b}
{Bloecker}, T. 1995{\natexlab{a}}, \aap, 299, 755

\bibitem[{{Bloecker}(1995{\natexlab{b}})}]{bloe95a}
---. 1995{\natexlab{b}}, \aap, 297, 727

\bibitem[{{Brandt}(2021)}]{bran21}
{Brandt}, T.~D. 2021, \apjs, 254, 42

\bibitem[{Brown {et~al.}(2011)Brown, Kilic, Brown, \& Kenyon}]{Brown2011}
Brown, J.~M., Kilic, M., Brown, W.~R., \& Kenyon, S.~J. 2011, \apj, 730, 67

\bibitem[{Brown {et~al.}(2016)Brown, Kilic, Kenyon, \& Gianninas}]{Brown2016}
Brown, W.~R., Kilic, M., Kenyon, S.~J., \& Gianninas, A. 2016, \apj, 824, 46

\bibitem[{{Brown} {et~al.}(2022){Brown}, {Kilic}, {Kosakowski}, \&
  {Gianninas}}]{Brown2022}
{Brown}, W.~R., {Kilic}, M., {Kosakowski}, A., \& {Gianninas}, A. 2022, \apj,
  933, 94

\bibitem[{{Brown} {et~al.}(2020){Brown}, {Kilic}, {Kosakowski}, {Andrews},
  {Heinke}, {Ag{\"u}eros}, {Camilo}, {Gianninas}, {Hermes}, \&
  {Kenyon}}]{brow20}
{Brown}, W.~R., {Kilic}, M., {Kosakowski}, A., {et~al.} 2020, \apj, 889, 49

\bibitem[{{Camisassa} {et~al.}(2019){Camisassa}, {Althaus}, {C{\'o}rsico}, {De
  Ger{\'o}nimo}, {Miller Bertolami}, {Novarino}, {Rohrmann}, {Wachlin}, \&
  {Garc{\'\i}a-Berro}}]{Camisassa2019}
{Camisassa}, M.~E., {Althaus}, L.~G., {C{\'o}rsico}, A.~H., {et~al.} 2019,
  \aap, 625, A87

\bibitem[{{Cheng} {et~al.}(2019){Cheng}, {Cummings}, \& {M{\'e}nard}}]{chen19}
{Cheng}, S., {Cummings}, J.~D., \& {M{\'e}nard}, B. 2019, \apj, 886, 100

\bibitem[{{Crumpler} {et~al.}(2024){Crumpler}, {Chandra}, {Zakamska}, {Adamane
  Pallathadka}, {Arseneau}, {Gentile Fusillo}, {Hermes}, {Badenes},
  {Chakraborty}, {G{\"a}nsicke}, \& {Schmidt}}]{Crumpler2024}
{Crumpler}, N.~R., {Chandra}, V., {Zakamska}, N.~L., {et~al.} 2024, \apj, 977,
  237

\bibitem[{{Cumming} {et~al.}(2008){Cumming}, {Butler}, {Marcy}, {Vogt},
  {Wright}, \& {Fischer}}]{cumm08}
{Cumming}, A., {Butler}, R.~P., {Marcy}, G.~W., {et~al.} 2008, \pasp, 120, 531

\bibitem[{{Cummings} {et~al.}(2018){Cummings}, {Kalirai}, {Tremblay},
  {Ramirez-Ruiz}, \& {Choi}}]{Cummings2018}
{Cummings}, J.~D., {Kalirai}, J.~S., {Tremblay}, P.~E., {Ramirez-Ruiz}, E., \&
  {Choi}, J. 2018, \apj, 866, 21

\bibitem[{{Decin} {et~al.}(2019){Decin}, {Homan}, {Danilovich}, {de Koter},
  {Engels}, {Waters}, {Muller}, {Gielen}, {Garc{\'\i}a-Hern{\'a}ndez},
  {Stancliffe}, {Van de Sande}, {Molenberghs}, {Kerschbaum}, {Zijlstra}, \& {El
  Mellah}}]{deci19}
{Decin}, L., {Homan}, W., {Danilovich}, T., {et~al.} 2019, Nature Astronomy, 3,
  408

\bibitem[{{Doherty} {et~al.}(2015){Doherty}, {Gil-Pons}, {Siess}, {Lattanzio},
  \& {Lau}}]{dohe15}
{Doherty}, C.~L., {Gil-Pons}, P., {Siess}, L., {Lattanzio}, J.~C., \& {Lau}, H.
  H.~B. 2015, \mnras, 446, 2599

\bibitem[{{Donada} {et~al.}(2023){Donada}, {Anders}, {Jordi}, {Masana},
  {Gieles}, {Perren}, {Balaguer-N{\'u}{\~n}ez}, {Castro-Ginard},
  {Cantat-Gaudin}, \& {Casamiquela}}]{dona23}
{Donada}, J., {Anders}, F., {Jordi}, C., {et~al.} 2023, \aap, 675, A89

\bibitem[{Duch{\^{e}}ne \& Kraus(2013)}]{Duchene2013}
Duch{\^{e}}ne, G., \& Kraus, A. 2013, \araa, 51, 269

\bibitem[{{Duquennoy} \& {Mayor}(1991)}]{duqu91}
{Duquennoy}, A., \& {Mayor}, M. 1991, \aap, 248, 485

\bibitem[{{El-Badry} \& {Rix}(2018)}]{elba18}
{El-Badry}, K., \& {Rix}, H.-W. 2018, \mnras, 480, 4884

\bibitem[{El-Badry {et~al.}(2021)El-Badry, Rix, \& Heintz}]{El-Badry2021}
El-Badry, K., Rix, H.-W., \& Heintz, T.~M. 2021, \mnras, 506, 2269

\bibitem[{El-Badry {et~al.}(2018)El-Badry, Ting, Rix, Quataert, Weisz, Cargile,
  Conroy, Hogg, Bergemann, \& Liu}]{El-Badry2018}
El-Badry, K., Ting, Y.-S., Rix, H.-W., {et~al.} 2018, \mnras, 476, 528

\bibitem[{Evans {et~al.}(2018)Evans, Riello, {De Angeli}, Carrasco,
  Montegriffo, Fabricius, Jordi, Palaversa, Diener, Busso, Cacciari, van
  Leeuwen, Burgess, Davidson, Harrison, Hodgkin, Pancino, Richards, Altavilla,
  Balaguer-N{\'{u}}{\~{n}}ez, Barstow, Bellazzini, Brown, Castellani, Cocozza,
  {De Luise}, Delgado, Ducourant, Galleti, Gilmore, Giuffrida, Holl, Kewley,
  Koposov, Marinoni, Marrese, Osborne, Piersimoni, Portell, Pulone, Ragaini,
  Sanna, Terrett, Walton, Wevers, \& Wyrzykowski}]{Evans2018}
Evans, D.~W., Riello, M., {De Angeli}, F., {et~al.} 2018, \aap, 616, A4

\bibitem[{{Fezenko} {et~al.}(2022){Fezenko}, {Hwang}, \&
  {Zakamska}}]{Fezenko2022}
{Fezenko}, G.~B., {Hwang}, H.-C., \& {Zakamska}, N.~L. 2022, \mnras, 511, 3881

\bibitem[{{Fischer} \& {Marcy}(1992)}]{fisc92}
{Fischer}, D.~A., \& {Marcy}, G.~W. 1992, \apj, 396, 178

\bibitem[{{Fontaine} {et~al.}(2001){Fontaine}, {Brassard}, \&
  {Bergeron}}]{font01}
{Fontaine}, G., {Brassard}, P., \& {Bergeron}, P. 2001, \pasp, 113, 409

\bibitem[{{Fregeau} {et~al.}(2009){Fregeau}, {Richer}, {Rasio}, \&
  {Hurley}}]{freg09}
{Fregeau}, J.~M., {Richer}, H.~B., {Rasio}, F.~A., \& {Hurley}, J.~R. 2009,
  \apjl, 695, L20

\bibitem[{{Gentile Fusillo} {et~al.}(2019){Gentile Fusillo}, Tremblay,
  G{\"{a}}nsicke, Manser, Cunningham, Cukanovaite, Hollands, Marsh, Raddi,
  Jordan, Toonen, Geier, Barstow, \& Cummings}]{GentileFusillo2019}
{Gentile Fusillo}, N.~P., Tremblay, P.-E., G{\"{a}}nsicke, B.~T., {et~al.}
  2019, \mnras, 482, 4570

\bibitem[{{Gentile Fusillo} {et~al.}(2021){Gentile Fusillo}, {Tremblay},
  {Cukanovaite}, {Vorontseva}, {Lallement}, {Hollands}, {G{\"a}nsicke},
  {Burdge}, {McCleery}, \& {Jordan}}]{gent21}
{Gentile Fusillo}, N.~P., {Tremblay}, P.~E., {Cukanovaite}, E., {et~al.} 2021,
  \mnras, 508, 3877

\bibitem[{{Goldman} {et~al.}(2017){Goldman}, {van Loon}, {Zijlstra}, {Green},
  {Wood}, {Nanni}, {Imai}, {Whitelock}, {Matsuura}, {Groenewegen}, \&
  {G{\'o}mez}}]{gold17}
{Goldman}, S.~R., {van Loon}, J.~T., {Zijlstra}, A.~A., {et~al.} 2017, \mnras,
  465, 403

\bibitem[{{Halbwachs} {et~al.}(2023){Halbwachs}, {Pourbaix}, {Arenou},
  {Galluccio}, {Guillout}, {Bauchet}, {Marchal}, {Sadowski}, \&
  {Teyssier}}]{halb23}
{Halbwachs}, J.-L., {Pourbaix}, D., {Arenou}, F., {et~al.} 2023, \aap, 674, A9

\bibitem[{{Hamer} \& {Schlaufman}(2019)}]{hame19}
{Hamer}, J.~H., \& {Schlaufman}, K.~C. 2019, \aj, 158, 190

\bibitem[{{Hamilton}(2022)}]{hami22}
{Hamilton}, C. 2022, \apjl, 929, L29

\bibitem[{{Hamilton} \& {Modak}(2024)}]{Hamilton2024}
{Hamilton}, C., \& {Modak}, S. 2024, \mnras, 532, 2425

\bibitem[{Hartman \& L{\'{e}}pine(2020)}]{Hartman2020}
Hartman, Z.~D., \& L{\'{e}}pine, S. 2020, \apjs, 247, 66

\bibitem[{{Heisler} \& {Tremaine}(1986)}]{heis86}
{Heisler}, J., \& {Tremaine}, S. 1986, \icarus, 65, 13

\bibitem[{{H{\"o}fner} \& {Olofsson}(2018)}]{hofn18}
{H{\"o}fner}, S., \& {Olofsson}, H. 2018, \aapr, 26, 1

\bibitem[{Hwang(2023)}]{Hwang2023}
Hwang, H.-C. 2023, \mnras, 518, 1750

\bibitem[{{Hwang} {et~al.}(2022{\natexlab{a}}){Hwang}, {El-Badry}, {Rix},
  {Hamilton}, {Ting}, \& {Zakamska}}]{Hwang2022twin}
{Hwang}, H.-C., {El-Badry}, K., {Rix}, H.-W., {et~al.} 2022{\natexlab{a}},
  \apjl, 933, L32

\bibitem[{{Hwang} {et~al.}(2024){Hwang}, {Ting}, {Cheng}, \&
  {Speagle}}]{Hwang2024}
{Hwang}, H.-C., {Ting}, Y.-S., {Cheng}, S., \& {Speagle}, J.~S. 2024, \mnras,
  528, 4272

\bibitem[{{Hwang} {et~al.}(2021){Hwang}, {Ting}, {Schlaufman}, {Zakamska}, \&
  {Wyse}}]{hwan21wide}
{Hwang}, H.-C., {Ting}, Y.-S., {Schlaufman}, K.~C., {Zakamska}, N.~L., \&
  {Wyse}, R. F.~G. 2021, \mnras, 501, 4329

\bibitem[{{Hwang} {et~al.}(2022{\natexlab{b}}){Hwang}, {Ting}, \&
  {Zakamska}}]{Hwang2022ecc}
{Hwang}, H.-C., {Ting}, Y.-S., \& {Zakamska}, N.~L. 2022{\natexlab{b}}, \mnras,
  512, 3383

\bibitem[{{Hwang} \& {Zakamska}(2020)}]{hwan20b}
{Hwang}, H.-C., \& {Zakamska}, N.~L. 2020, \mnras, 493, 2271

\bibitem[{{Iben}(1990)}]{Iben1990}
{Iben}, Icko, J. 1990, \apj, 353, 215

\bibitem[{{Ivanova} {et~al.}(2013{\natexlab{a}}){Ivanova}, {Justham}, {Avendano
  Nandez}, \& {Lombardi}}]{ivan13b}
{Ivanova}, N., {Justham}, S., {Avendano Nandez}, J.~L., \& {Lombardi}, J.~C.
  2013{\natexlab{a}}, Science, 339, 433

\bibitem[{{Ivanova} {et~al.}(2013{\natexlab{b}}){Ivanova}, {Justham}, {Chen},
  {De Marco}, {Fryer}, {Gaburov}, {Ge}, {Glebbeek}, {Han}, {Li}, {Lu}, {Marsh},
  {Podsiadlowski}, {Potter}, {Soker}, {Taam}, {Tauris}, {van den Heuvel}, \&
  {Webbink}}]{ivan13a}
{Ivanova}, N., {Justham}, S., {Chen}, X., {et~al.} 2013{\natexlab{b}}, \aapr,
  21, 59

\bibitem[{{Izzard} {et~al.}(2010){Izzard}, {Dermine}, \& {Church}}]{izza10}
{Izzard}, R.~G., {Dermine}, T., \& {Church}, R.~P. 2010, \aap, 523, A10

\bibitem[{{Jiang} \& {Tremaine}(2010)}]{jian10}
{Jiang}, Y.-F., \& {Tremaine}, S. 2010, \mnras, 401, 977

\bibitem[{{Kilic} {et~al.}(2018){Kilic}, {Hambly}, {Bergeron},
  {Genest-Beaulieu}, \& {Rowell}}]{kili18}
{Kilic}, M., {Hambly}, N.~C., {Bergeron}, P., {Genest-Beaulieu}, C., \&
  {Rowell}, N. 2018, \mnras, 479, L113

\bibitem[{{Klein} \& {Katz}(2017)}]{klei17}
{Klein}, Y.~Y., \& {Katz}, B. 2017, \mnras, 465, L44

\bibitem[{Kozai(1962)}]{Kozai1962}
Kozai, Y. 1962, \aj, 67, 591

\bibitem[{{Kroupa}(2001)}]{krou01}
{Kroupa}, P. 2001, \mnras, 322, 231

\bibitem[{{Kruckow} {et~al.}(2021){Kruckow}, {Neunteufel}, {Di Stefano}, {Gao},
  \& {Kobayashi}}]{Kruckow2021}
{Kruckow}, M.~U., {Neunteufel}, P.~G., {Di Stefano}, R., {Gao}, Y., \&
  {Kobayashi}, C. 2021, \apj, 920, 86

\bibitem[{{Lamers} \& {Levesque}(2017)}]{lame17}
{Lamers}, H. J.~G.~L.~M., \& {Levesque}, E.~M. 2017, {Understanding Stellar
  Evolution} (Bristol, UK: IOP Publishing), doi:10.1088/978-0-7503-1278-3

\bibitem[{Law {et~al.}(2010)Law, Dhital, Kraus, Stassun, \& West}]{Law2010}
Law, N.~M., Dhital, S., Kraus, A.~L., Stassun, K.~G., \& West, A.~A. 2010,
  \apj, 720, 1727

\bibitem[{Lidov(1962)}]{Lidov1962}
Lidov, M. 1962, Planetary and Space Science, 9, 719

\bibitem[{Lindegren {et~al.}(2018)Lindegren, Hernandez, Bombrun, Klioner,
  Bastian, Ramos-Lerate, de~Torres, Steidelmuller, Stephenson, Hobbs, Lammers,
  Biermann, Geyer, Hilger, Michalik, Stampa, McMillan, Castaneda, Clotet,
  Comoretto, Davidson, Fabricius, Gracia, Hambly, Hutton, Mora, Portell, van
  Leeuwen, Abbas, Abreu, Altmann, Andrei, Anglada, Balaguer-Nunez, Barache,
  Becciani, Bertone, Bianchi, Bouquillon, Bourda, Brusemeister, Bucciarelli,
  Busonero, Buzzi, Cancelliere, Carlucci, Charlot, Cheek, Crosta, Crowley,
  de~Bruijne, de~Felice, Drimmel, Esquej, Fienga, Fraile, Gai, Garralda,
  Gonzalez-Vidal, Guerra, Hauser, Hofmann, Holl, Jordan, Lattanzi, Lenhardt,
  Liao, Licata, Lister, Loffler, Marchant, Martin-Fleitas, Messineo, Mignard,
  Morbidelli, Poggio, Riva, Rowell, Salguero, Sarasso, Sciacca, Siddiqui,
  Smart, Spagna, Steele, Taris, Torra, van Elteren, van Reeven, \&
  Vecchiato}]{Lindegren2018}
Lindegren, L., Hernandez, J., Bombrun, A., {et~al.} 2018, \aap, 616, A2

\bibitem[{{Maxted} {et~al.}(2006){Maxted}, {Napiwotzki}, {Dobbie}, \&
  {Burleigh}}]{Maxted2006}
{Maxted}, P.~F.~L., {Napiwotzki}, R., {Dobbie}, P.~D., \& {Burleigh}, M.~R.
  2006, \nat, 442, 543

\bibitem[{{Miller Bertolami}(2016)}]{mill16}
{Miller Bertolami}, M.~M. 2016, \aap, 588, A25

\bibitem[{{Modak} \& {Hamilton}(2023)}]{moda23}
{Modak}, S., \& {Hamilton}, C. 2023, \mnras, 524, 3102

\bibitem[{Moe \& {Di Stefano}(2017)}]{Moe2017}
Moe, M., \& {Di Stefano}, R. 2017, \apjs, 230, 15

\bibitem[{{Murray} \& {Dermott}(1999)}]{murr99}
{Murray}, C.~D., \& {Dermott}, S.~F. 1999, {Solar system dynamics} (Cambridge
  Univ. Press)

\bibitem[{{O'Connor} {et~al.}(2023){O'Connor}, {Lai}, \&
  {Seligman}}]{OConnor2023}
{O'Connor}, C.~E., {Lai}, D., \& {Seligman}, D.~Z. 2023, \mnras, 524, 6181

\bibitem[{{Paczy{\'n}ski}(1971)}]{pacz71}
{Paczy{\'n}ski}, B. 1971, \araa, 9, 183

\bibitem[{{Pham} \& {Rein}(2024)}]{Pham2024}
{Pham}, D., \& {Rein}, H. 2024, \mnras, 530, 2526

\bibitem[{{Raghavan} {et~al.}(2010){Raghavan}, {McAlister}, {Henry}, {Latham},
  {Marcy}, {Mason}, {Gies}, {White}, \& {ten Brummelaar}}]{ragh10}
{Raghavan}, D., {McAlister}, H.~A., {Henry}, T.~J., {et~al.} 2010, \apjs, 190,
  1

\bibitem[{{Rein} \& {Liu}(2012)}]{rein12}
{Rein}, H., \& {Liu}, S.-F. 2012, \aap, 537, A128

\bibitem[{{Renedo} {et~al.}(2010){Renedo}, {Althaus}, {Miller Bertolami},
  {Romero}, {C{\'o}rsico}, {Rohrmann}, \& {Garc{\'\i}a-Berro}}]{rene10}
{Renedo}, I., {Althaus}, L.~G., {Miller Bertolami}, M.~M., {et~al.} 2010, \apj,
  717, 183

\bibitem[{{Steffen} {et~al.}(1998){Steffen}, {Szczerba}, \&
  {Schoenberner}}]{stef98}
{Steffen}, M., {Szczerba}, R., \& {Schoenberner}, D. 1998, \aap, 337, 149

\bibitem[{{Tokovinin}(2014)}]{toko14}
{Tokovinin}, A. 2014, \aj, 147, 87

\bibitem[{Tokovinin(2017)}]{Tokovinin2017}
Tokovinin, A. 2017, \mnras, 468, 3461

\bibitem[{{Tokovinin} {et~al.}(2006){Tokovinin}, {Thomas}, {Sterzik}, \&
  {Udry}}]{toko06}
{Tokovinin}, A., {Thomas}, S., {Sterzik}, M., \& {Udry}, S. 2006, \aap, 450,
  681

\bibitem[{Tokovinin(1998)}]{Tokovinin1998}
Tokovinin, A.~A. 1998, Astronomy Letters, 24, 178

\bibitem[{{van Roestel} {et~al.}(2021){van Roestel}, {Kupfer}, {Bell},
  {Burdge}, {Mr{\'o}z}, {Prince}, {Bellm}, {Drake}, {Dekany}, {Mahabal},
  {Porter}, {Riddle}, {Shin}, {Shupe}, \& {Kulkarni}}]{VanRoestel2021}
{van Roestel}, J., {Kupfer}, T., {Bell}, K.~J., {et~al.} 2021, \apjl, 919, L26

\bibitem[{{Vassiliadis} \& {Wood}(1994)}]{vass94}
{Vassiliadis}, E., \& {Wood}, P.~R. 1994, \apjs, 92, 125

\bibitem[{{Weiss} \& {Ferguson}(2009)}]{weis09}
{Weiss}, A., \& {Ferguson}, J.~W. 2009, \aap, 508, 1343

\bibitem[{{Xu} {et~al.}(2023){Xu}, {Hwang}, {Hamilton}, \& {Lai}}]{xu23}
{Xu}, S., {Hwang}, H.-C., {Hamilton}, C., \& {Lai}, D. 2023, \apjl, 949, L28

\bibitem[{{Zakamska} \& {Hwang}(2025)}]{zenodo2025}
{Zakamska}, N., \& {Hwang}, H.-C. 2025, Software for computing the effects of
  stellar evolution and mass loss on orbits of wide stellar binaries,
  \\url{https://zenodo.org/record/15991774}, v.1.0,  Zenodo,
  doi:10.5281/zenodo.15991774

\end{thebibliography}
\bibliographystyle{aasjournal}

\end{document}